# XUV exposed, non-hydrostatic hydrogen-rich upper atmospheres of terrestrial planets II: Hydrogen coronae and ion escape

Kristina G. Kislyakova[1,2], Helmut Lammer[1], Mats Holmström[3], Mykhaylo Panchenko[1],
Petra Odert[1,2], Nikolai V. Erkaev[4], Martin Leitzinger[2],
Maxim L. Khodachenko[1], Yuri N. Kulikov[5],
Manuel Güdel[6], Arnold Hanslmeier[2]

[1]Austrian Academy of Sciences, Space Research Institute,
Schmiedlstr. 6, A-8042 Graz, Austria
(kristina.kislyakova@oeaw.ac.at, maxim.khodachenko@oeaw.ac.at,
helmut.lammer@oeaw.ac.at, petra.odert@oeaw.ac.at, mykhaylo.panchenko@oeaw.ac.at)
[2]Institute for Physics, University of Graz, Universitätsplatz 5, A-8010 Graz, Austria
(martin.leitzinger@uni-graz.at, arnold.hanslmeier@uni-graz.at)
[3]Swedish Institute of Space Physics, P.O. Box 812, SE-98128 Kiruna, Sweden
(matsh@irf.se)
[4]Institute of Computational Modelling, Siberian Division of Russian Academy of
Sciences, Akademgorodok 28/44 660036 Krasnoyarsk, Russian Federation
(erkaev@icm.krasn.ru)
[5]Polar Geophysical Institute (PGI), Russian Academy of Sciences,
Khalturina Str. 15, Murmansk, 183010, Russian Federation
(kulikov@pgi.ru)
[6]Institute for Astrophysics, University of Vienna, Türkenschanzstr. 17, 1180, Austria
(manuel.guedel@univie.ac.at)

Corresponding Authors:
Kristina G. Kislyakova
E-mail: kislyakova.kristina@oeaw.ac.at
Austrian Academy of Sciences
Space Research Institute
Schmiedlstr. 6, A-8042 Graz
Austria

**ABSTRACT**

We study the interactions between the stellar wind plasma flow of a typical M star, such as GJ 436, and hydrogen-rich upper atmospheres of an Earth-like planet and a "super-Earth" with the radius of $2R_{Earth}$ and a mass of $10M_{Earth}$, located within the habitable zone at ~0.24 AU. We investigate the formation of extended atomic hydrogen coronae under the influences of the stellar XUV flux (soft X-rays and EUV), stellar wind density and velocity, shape of a planetary obstacle (e.g., magnetosphere, ionopause), and the loss of planetary pick-up ions on the evolution of hydrogen-dominated upper atmospheres. Stellar XUV fluxes which are 1, 10, 50 and 100 times higher compared to that of the present-day Sun are considered and the formation of high-energy neutral hydrogen clouds around the planets due to the charge-exchange reaction under various stellar conditions have been modeled. Charge-exchange between stellar wind protons with planetary hydrogen atoms, and photoionization, leads to the production of initially cold ions of planetary origin. We found that the ion production rates for the studied planets can vary over a wide range, from ~$1.0 \times 10^{25} \, s^{-1}$ to ~$5.3 \times 10^{30} \, s^{-1}$, depending on the stellar wind conditions and the assumed XUV exposure of the upper atmosphere. Our findings indicate that most likely the majority of these planetary ions are picked up by the stellar wind and lost from the planet. Finally, we estimate the long-time non-thermal ion pick-up escape for the studied planets and compare them with the thermal escape. According to our estimates, non-thermal escape of picked up ionized hydrogen atoms over a planet's lifetime varies between ~0.4 Earth ocean equivalent amounts of hydrogen ($EO_H$) to < 3 $EO_H$ and usually is several times smaller in comparison to the thermal atmospheric escape rates.

Keywords: stellar activity, low mass stars, early atmospheres, Earth-like exoplanets, ENAs, ion escape, habitability

## 1. INTRODUCTION

Recent discoveries of so-called low density "super-Earths" by various ground- and space-based exoplanet-transit surveys indicate large populations of volatile-rich big rocky planets. Findings from ESOs High Accuracy Radial velocity Planetary Search project (HARPS), and from NASAs Kepler space observatory, revealed that planets which are slightly larger and more massive compared to the Earth may be very common in the Universe. From the available statistics and the discovery of Kepler-22b, a "super-Earth" with the size of about $2.38 \pm 0.13 R_{Earth}$ within the habitable zone (HZ) of a Sun-type star, one can expect that planets orbiting within the HZ should be frequent in the Universe and should also orbit cooler, lower mass M dwarfs. Approximately 100 of them are found in the immediate neighborhood of the Sun (Scalo et al. 2007; Bonfils et al. 2011). These earlier estimations are now supported by a recent study by Dressing and Charbonneau (2013), which used optical and near-infrared photometry from the Kepler Input Catalog to estimate the occurrence rate of Earth-like planets orbiting dwarf stars. The estimation of Dressing and Charbonneau (2013), from the 248 early M dwarfs within 10 parsecs of the Sun, shows that there should be at least 9 Earth-size planets in their habitable zones.

Moreover, from the radius-mass relation and the resulting density of discovered "super-Earths", one finds that these bodies probably have rocky cores but are surrounded by significant H/He and/or $H_2O$ envelopes. These findings are in agreement with recent theoretical studies, which suggest that small planets are not necessarily rocky Earth-like bodies (e.g., Wuchterl 1993; Kuchner 2003; Léger $et$ $al.$ 2004; Ikoma and Hory 2012; Elkins-Tanton 2011; Lammer, 2012; Lammer $et$ $al.$ 2011a). For explaining the mean density of Kepler 11d, Kepler 11e, and Kepler 11f these "super-Earths'' require dense H/He envelopes, similar to Uranus and Neptune, while Kepler-11b and 11c may have also additional $H_2O$ to their H/He gas envelopes (Lissauer $et$ $al.$ 2011), and GJ 1214b (Charbonneau $et$ $al.$, 2009) or 55Cnc e (Endl $et$ $al.$, 2012) may contain a huge amount of $H_2O$.

If Earth-like and "super-Earth"- type exoplanets can accumulate hydrogen from the nebula gas of an equivalent amount of 100 to 1000, and even up to $10^4$ times, that of an Earth ocean depends on the nebula dissipation time, the formation time of the protoplanet, its luminosity, and nebula characteristics such as grain depletion factors, etc. (e.g., Mizuno $et$ $al.$, 1978; Hayashi $et$ $al.$, 1979; Ikoma and Genda, 2006; Rafikov, 2006). Although Solar System planets such as Venus, Earth and Mars lost their nebula-based hydrogen envelopes during the first 100 Myr after their origin, or never accumulated such huge amounts due to step-wise accretion after the nebula gas disappeared, terrestrial planets in other systems evolve under different conditions and may capture such a dense protoatmosphere which they may not lose during the extreme active period of their host stars.

To understand how frequent "rocky" terrestrial planets really are, more observations are certainly needed. From the available statistics one can conclude that Earth-analogue class I habitats (Lammer *et al.*, 2009a; Lammer, 2013) have to be

- located at the right distance inside the HZ of their host stars,
- must lose their nebula-captured H/He or degassed $H_2O$ and volatile-rich proto-atmospheres during the right time period, i.e. not remain as mini-Neptune type bodies,
- should maintain plate tectonics, liquid water and landmass above the water level over the planet's lifetime,

and

- nitrogen should be the main atmospheric species after the stellar activity decreased to moderate values.

The question if more massive "super-Earths" can maintain plate tectonics over time spans of several Gyr is controversial (e.g., Valencia *et al.*, 2007; van Heck and Tackley, 2011; Korenaga, 2010). However, in this work we will not discuss the pro and contra about geophysical processes, but focus on the stellar wind erosion of captured H/He envelopes, orand? the hydrogen content of outgassed hydrogen-rich steam atmospheres, because the proto-atmosphere escape determines if a planet will evolve to an Earth-like habitat or may remain as a mini-Neptune.

As it is shown by Erkaev *et al.* (2013) (part I of this study), depending on the availability of possible IR-cooling molecules and the planets average density, hydrogen-rich "super-Earths" orbiting inside the HZ will experience hydrodynamic blow-off only for XUV fluxes several 10 times higher compared to today's Sun. Most of their lifetime the upper atmospheres of these planets will experience strong Jeans escape which is still weaker compared to blow-off, so that they may not lose efficiently their dense hydrogen envelopes. Jeans escape is the classical thermal escape mechanism based on the fact that the atmospheric particles have velocities according to the Maxwell distribution. Individual particles in the high tail of the distribution may reach escape velocity at the exobase altitude, where the mean free path is comparable to the scale height, so that they can escape from the planet's atmosphere. When the thermosphere temperature rises due to heating by the stellar XUV radiation, the number of these energetic particles increases and the atmosphere finally reaches the state when the majority of the particles have velocities equal to or exceeding the escape velocity. In this case the atmosphere is not hydrostatic anymore, and starts to expand similar to the Parker-type solar corona. This mechanism is called blow-off and leads to a stronger escape in comparison to the Jeans mechanism.

The blow-off stage is more easily reached at less massive hydrogen-rich planets with mass equal to that of the Earth. These planets experience hydrodynamic blow-off for much longer, and change from the blow-off regime to the Jeans-type escape for XUV fluxes which are < 10 times of today's Sun. Because of XUV heating and expansion of their upper atmospheres, both of our test planets should produce extended exospheres or hydrogen coronae distributed above possible magnetic obstacles defined by intrinsic or induced magnetic fields. In such cases the hydrogen-rich upper atmosphere will not be protected by possible magnetospheres like on present-day Earth, but could be eroded by the stellar wind plasma flow and lost from the planet in the form of ions (Erkaev *et al.*, 2005; Lammer *et al.*, 2007).

Besides thermal escape from the hydrogen-dominated upper atmosphere of the two considered test-planets (Erkaev *et al.*, 2013), briefly discussed above, one can expect that non-thermal atmospheric escape processes will also contribute to the losses. Non-thermal escape processes can be separated in ion escape and photochemical, as well as kinetic, processes which accelerate atoms beyond escape energy. Ions can escape from an upper atmosphere if the exosphere is not protected by a strong magnetic field and stretches above the magnetopause. In such a case exospheric neutral atoms can interact with the host stars solar/stellar plasma (i.e., winds, CMEs) environment. The hydrogen atoms which flow upward from the lower thermosphere will be ionized by the stellar radiation, electron impact or charge exchange and then accelerated by electric fields within the solar/stellar wind plasma flow around the planetary obstacle (i.e. ionopause, magnetopause), so that they are finally picked up and lost form the planet's gravity field (e.g., Lammer *et al.*, 2007; Ma and Nagy, 2007; Lammer, 2013).

From space missions to non- or weakly magnetized planets such as Venus and/or Mars it is known that planetary ions can also be detached from an ionopause by plasma instabilities in the form of ionospheric clouds (Terada *et al*., 2002; Penz *et al.*, 2004; Möstl *et al*., 2011), or by momentum transport triggered outflow through the planetary tail. On Earth ions outflow also over Polar Regions along open magnetic field lines (Lundin *et al*., 2007; Yau and André, 1997; Wei *et al*., 2012).

From the analysis of the available ion escape data from Venus and Mars by the ASPERA instruments on board of Venus Express and Mars Express, as well as from theoretical models, one can conclude that ion pick-up is a very dominant permanently acting non-thermal ion escape process, and most likely more efficient compared to the sporadic losses triggered by plasma instabilities or outflow through the planets tail. However, there may be extreme solar events which can enhance the ion outflow sporadically by cool ion outflow or plasma instabilities.

Non-thermal escape processes of neutral atoms are caused by sputtering of atmospheric neutral atoms, photochemical processes such as dissociative recombination, and charge exchange. Direct escape by sputtering is only a relevant process for low mass bodies which have a mass ≤ Mars. utHowever, even in the case of Mars one can expect that sputter loss rates are an order of magnitude lower compared for instance to ion pick-up (e.g., Leblanc and Johnson, 2002; Chassefière and Leblanc, 2004; Lammer *et al*., 2013). Direct escape of heavy neutral atoms, such as O and C, caused by photochemical processes is expected to be higher compared to ion escape from Mars (e.g., Krestyanikova and Shematovich, 2005; Chaufray et al., 2007; 2006; Fox and Haċ, 2009; Lammer et al., 2013) but negligible or lower at more massive planets such as Venus (Gröller et al., 2010; 2012) or the Earth.

Lighter atoms such as atomic hydrogen can also escape directly from more massive planets with escape rates lower or comparable to ion escape. Theoretical models which studied the photochemical escape rates of H atoms (Shematovich, 2010) and ion pick-up ion escape rates (Erkaev *et al*., 2005) from the hot Jupiter HD 208459b indicate comparable loss rates of the order of $\leq 10^9$ g s$^{-1}$, which are an order of magnitude lower compared to the modeled thermal escape (e.g., Yelle, 2004; 2006; Tian *et al*., 2005a;

Penz *et al*., 2008b; García Muñoz, 2007; Murray-Clay *et al.*, 2009; Linsky *et al*., 2010; Koskinen *et al*., 2012).

From the brief overview on various non-thermal atmospheric escape processes, one can conclude that stellar wind induced ion erosion from XUV-heated and extended hydrogen-rich thermospheres (Erkaev *et al*., 2013), where H atoms will most likely not be protected by a possible magnetosphere, so-called ion pick-up will be one of the most efficient non-thermal atmospheric escape process. Because of many unknowns related to minor atmospheric species in exoplanet atmospheres, as well as magnetic field properties, a study of more complex but most likely less effective processes, such as cool ion or polar outflow and photochemical non-thermal escape processes which are not even well understood at Solar System planets including the Earth, would yield highly speculative results. Therefore, in this study we focus on the modeling of the stellar wind plasma interaction, related ion production rates via charge-exchange and photoionization, and escape estimates of planetary pick-up ions from XUV exposed upper atmospheres which originate from hydrogen-rich thermospheres of an Earth-like ($R_{pl}$=1$R_{Earth}$, $M_{pl}$=1$M_{Earth}$) planet in comparison with a "super-Earth" ($R_{pl}$=2$R_{Earth}$, $M_{pl}$=10$M_{Earth}$). For reasons of comparative escape studies between thermal and non-thermal ion pick-up, we study the same test planets as investigated by Erkaev *et al.* (2013) within an orbit of a typical HZ of an M star with the size and mass of ~0.45$R_{Sun}$. For the host star of our test planets we use the well observed dwarf star GJ 436 (Ehrenreich *et al.*, 2011; von Braun *et al.*, 2012, France *et al.*, 2012) as a proxy.

So far the formation of such extended hydrogen coronae was only addressed in a brief way in Lammer *et al.* (2011a; 2011b), but never modeled in detail. In this study we apply a coupled Direct Simulation Monte Carlo (DSMC) upper atmosphere - stellar wind plasma interaction model (Holmström *et al.*, 2008; Ekenbäck *et al.*, 2010) to the results of Erkaev *et al.* (2012).

In Sect. 2 we describe the DSMC model, which is used for the calculation of the exosphere and related hydrogen coronae, as well as the coupled solar/stellar wind plasma upper atmosphere interaction model. We validate our model by applying it to the Earth's geocorona and comparing the simulation results with the present-day exosphere hydrogen density and energetic neutral atom (ENA) observations by NASAs Interstellar Boundary

Explorer (IBEX) satellite near the magnetopause boundary at ~10 $R_{Earth}$. After validating our model for the geocorona of present-day Earth, in Sect. 3 we describe the radiation and plasma parameters of our chosen M-type host star proxy, Gliese 436. In Sect. 4 we present the modeling results for the extended hydrogen coronae. In Sect. 4.1 the results of the stellar wind plasma interaction and the production of ENAs are shown as well as related planetary hydrogen ion pick-up escape rates as a function of the XUV flux values from 1 to 100 times that of today's Sun. The atmospheric ion escape rates are compared with the thermal hydrogen neutral loss rates modeled by Erkaev *et al.* (2012) in Section 4.2. In Section 4.3 we estimate the possible mass loss of hydrogen ions during the planetary lifetime and discuss the implications of our findings for the evolution of Earth-like and more massive "super-Earths". Section 5 summarizes the findings of our study.

## 2 STELLAR WIND-UPPER ATMOSPHERE INTERACTION

As shown by previous studies of Watson *et al.* (1981), Kasting and Pollack (1983), Tian *et al.* (2005b; 2008a; 2008b), Volkov *et al.* (2011), and Erkaev *et al.* (2013), hydrogen-rich terrestrial planets experience XUV-heated and hydrodynamically expanding non-hydrostatic upper atmosphere conditions. Depending on the particular environment (e.g., XUV flux, orbital distance, availability of IR-cooling molecules, the planet's average density, etc.) the results of these studies indicate that such planets can expand their exobase level, which separates the collision dominated atmosphere from the collisionless region to distances from a few $R_{pl}$ up to more than $20R_{pl}$. As a result of such an expansion of the upper atmosphere an intrinsic planetary magnetic field will most likely not protect the exosphere against the stellar wind plasma flow (Lichtenegger *et al.*, 2010; Lammer, 2012; Lammer *et al.*, 2011a).

Due to the interaction between the stellar wind plasma flow and the XUV-heated non-hydrostatic upper neutral atmosphere of the planet, energetic neutral atoms (ENAs) are produced. ENAs originate due to charge-exchange when an electron is transferred from a planetary neutral atom to a stellar wind proton which then becomes an ENA. This interaction process between the stellar wind plasma and the upper atmosphere plays a significant role in the ion erosion of upper planetary atmospheres (e.g., Lundin *et al.*, 2007; Lammer, 2013). The production of ENAs after the interaction of stellar wind protons via charge exchange with various upper atmosphere species is described by the reactions in eq. (1-3).

$$H^+_{SW} + H_{pl} = H^+_{pl} + H_{ENA} \qquad (1)$$

$$H^+_{SW} + O_{pl} = O^+_{pl} + H_{ENA} \qquad (2)$$

$$H^+_{SW} + N_{pl} = N^+_{pl} + H_{ENA} \qquad (3)$$

After its production an ENA continues to travel with the initial velocity and energy of the stellar wind proton. The atmospheric atom in turn becomes an initially cold ion which can afterwards be lost from the atmosphere due to the ion pick-up process (Lammer, 2013). In the current study we focus our attention only on the reaction shown in eq. (1) which will dominate the stellar wind interaction with planetary hydrogen coronae around hydrogen-rich terrestrial planets.

## 2.1 Model description

In the current study the plasma interaction between the stellar wind and the upper atmosphere of the Earth-like planet and "super-Earth" is modeled applying a Direct Simulation Monte Carlo (DSMC) upper atmosphere-exosphere particle model which is coupled with a stellar wind particle interaction code. The 3D model is described in detail in Holmström *et al.* (2008) and Ekenbäck *et al.* (2010) and includes stellar wind protons and planetary hydrogen atoms. The latter are launched into the simulation domain from the upper atmosphere. The applied collision cross sections for hydrogen atoms, $\sigma_{\text{H-H}}$, and for protons and hydrogen atoms, $\sigma_{\text{H+-H}}$, are $10^{-17}$ cm$^2$ and $2 \times 10^{-15}$ cm$^2$ (Ekenbäck *et al.*, 2010) respectively. Charge exchange between stellar wind protons and exospheric hydrogen atoms takes place outside a conic shaped obstacle that represents the magneto-ionopause of the studied planet. Stellar wind protons that have charge-exchanged according to the reaction shown in eq. (1) become ENAs.

Besides of the charge-exchange reaction, the model includes gravitation of the planet and tidal effects as well as scattering by atmospheric atoms of UV photons (radiation pressure) and photoionization by stellar photons. Inclusion of the tidal-generating potential into the equations leads to the extension of the atmosphere toward and backward from the host star, and in extreme cases to Roche lobe overflow. Nevertheless, these effects are important for "hot Jupiters" which are located at very close distance to their host stars, and do not play a significant role for the test planets we consider in the present study. All collisions are modeled using a DSMC algorithm (Holmström *et al.,* 2008; Ekenbäck *et al.,* 2010). The main code uses the FLASH software developed at the University of Chicago which provides adaptive grids and is fully parallelized (Fryxell *et al.,* 2000). The coordinate system is centered at the center of the planet with mass $M_{pl}$, the $x_1$-axis is pointing towards the center of mass of the system, the $x_3$-axis is parallel to the direction of the angular velocity of rotation $\Omega$, and the $x_2$-axis points in the opposite direction to the planet's velocity. $M_{St}$ is the mass of the planets host star.

Tidal potential, Coriolis and centrifugal forces, as well as the gravitation of the star and planet acting on a hydrogen neutral atom, are included in the code in the following way (Chandrasekhar, 1963)

$$\frac{dv_i}{dt} = \frac{\partial}{\partial x_i}\left[\frac{1}{2}\Omega^2\left(x_1^2+x_2^2\right)+\mu\left(x_1^2-\frac{1}{2}x_2^2-\frac{1}{2}x_3^2\right)+\left(\frac{GM_{St}}{R^2}-\frac{M_{St}R}{M_{pl}+M_{St}}\Omega^2\right)x_1\right]+2\Omega\varepsilon_{il3}v_l \quad (4)$$

Here $v_i$ are the components of the velocity vector of a particle, $G$ is Newton's gravitational constant, $R$ the distance between the centers of mass, $\varepsilon$ the Levi-Civita symbol, and $\mu=\frac{GM_{St}}{R^3}$. The first term in the right-hand side of the eq. (4) represents the centrifugal force, the second is the tidal-generating potential, the third the gravitation of the planet's host star and the planet while the last term stands for the Coriolis force. The self-gravitational potential of a particle is neglected.

Charge exchange reactions between a neutral planetary hydrogen atom and a stellar wind proton may take place outside the obstacle representing a magneto- or ionopause

$$x_1 = R_s\left(1-\frac{x_2^2+x_3^2}{R_t^2}\right) \qquad (5)$$

Here $R_s$ stands for the magnetosphere or planetary obstacle stand-off distance and $R_t$ the width of the obstacle. Since the obstacle shape and location depend strongly on the planetary magnetic field strength, one may model the interaction of the stellar wind with magnetized as well as with non- or weakly magnetized planets by the appropriate choice of $R_s$ and $R_t$.

### 2.2 Exosphere modeling of Earth's observed atomic hydrogen geocorona

ENAs have been observed around all Solar System planets where a spacecraft was equipped with a corresponding instrument (e.g., Futaana *et al.*, 2006; Galli *et al.*, 2008; Lammer *et al.*, 2011a; 2011b). As shown in Fig. 1, the Interstellar Boundary Explorer (IBEX) satellite recently observed an ENA formation zone around Earth's subsolar magnetopause stand-off distance, which is located at 10 $R_{Earth}$ from the planet's center (Fuselier *et al.*, 2010).

Before we apply our model to hydrogen-rich exoplanets, we validate it by reproducing the geocorona and recent ENA observations (Fuselier *et al.* 2010) around the Earth's magnetopause by the NASAs IBEX satellite of present-day Earth. Fig. 2 shows our modeling results for Earth's geocorona interacting with the present-day solar wind by taking all parameters of Earth's exosphere as given in Table 1 as an input.

The average neutral hydrogen atom density at the magnetopause level obtained from our model is estimated to be ~8 cm$^{-3}$ at the distance of ~10$R_{Earth}$. This density value coincides very well with the exospheric number densities inferred from the IBEX observations of ENAs near the magnetopause. In the case of the IBEX observation at March 28 2009, the computed and observed proton fluxes show an exosphere hydrogen density at a geocentric distance of ~10$R_{Earth}$ of ~$4 - 11$ cm$^{-3}$ (Fuselier *et al.*, 2010). The estimates of the modeled ENA flux are in good agreement with the observed flux as well, predicting the flux of approximately 600 (cm² s sr keV)$^{-1}$. This value falls inside the observed ENA interval of ~$530 - 2300$ (cm² s sr keV)$^{-1}$ (Fuselier *et al.*, 2010).

The IBEX observation and our model validation can also be seen as a confirmation that under extreme radiation and plasma environments of the young Sun or more active stars a huge ENA formation zone, as suggested by Chassefière (1996) and Lammer *et al.* (2011; 2012), should be produced in the stellar wind interaction region of a hydrogen-rich extended upper atmosphere of an Earth-size planet when the exosphere density near the magnetopause is more than $10^6$ times larger than that observed at present-day Earth. In the following sections we describe the input parameters and our applied exosphere and ENA models to the XUV exposed hydrogen-rich Earth-size test planets.

### 3. GLIESE 436: A HOST STAR PROXY FOR HYDROGEN-RICH TERRESTRIAL TEST PLANETS

*3.1 The radiation environment of Gliese 436*

Recently Ehrenreich *et al.* (2011) observed with the Hubble Space Telescope Imaging Spectrograph (HST/STIS) the Lyman-α emission (1215.67 Å) of neutral hydrogen atoms from the low mass M star, GJ 436. Because this emission is a main contributor to the ultraviolet flux it can also be used as a main tracer in studies of thermospheric heating, thermal escape, and possible absorption by extended hydrogen coronae and/or ENAs (e.g. Vidal-Madjar *et al.*, 2003; Holmström *et al*., 2008; Ekenbäck *et al*., 2010; Ben-Jaffel and Hosseini, 2010; Lecavelier des Etangs *et al*., 2010; Lammer *et al.*, 2011b; Ehrenreich *et al*., 2012, Lammer, 2013) during transit observations with ultraviolet transmission spectroscopy. We use GJ 436 as a typical M-type host star for our test-planet parameter studies. GJ 436 is a M2.5 dwarf star which is 10.2 pc away from the Sun. The dwarf star hosts a transiting "hot Neptune" at an orbital distance of about 0.03 AU (Butler *et al.*, 2004; Gillon *et al.*, 2007). We adopt values for stellar mass and radius of 0.45 $M_{Sun}$ and 0.45 $R_{Sun}$, respectively, which are consistent with several independent parameter determinations of GJ 436 (Torres, 2007; Maness *et al.*, 2007; von Braun *et al.*, 2012). The location of the habitable zone (HZ) is calculated following Selsis *et al.* (2007). As their relations are only valid for effective temperatures down to 3700 K, this value is used instead of the true temperature of GJ 436, which is slightly lower (3400-3600 K; Torres 2007; von Braun *et al.* 2012). Further, we adopt a bolometric luminosity of 0.026 $L_{Sun}$ (Torres, 2007). This leads to a HZ extent of 0.12-0.36 AU assuming the limit of 50% clouds, as typical for the Earth. Hence, the center of the HZ is located at 0.24 AU, which we adopt as the orbit of our hypothetical exo-Earth. The orbital period of an Earth-analog planet within the HZ of GJ 436 corresponds to approximately 63.7 days, the orbital velocity to about 41 km s$^{-1}$, and the angular velocity to 1.14·10$^{-6}$ rad s$^{-1}$. The age of GJ 436 is about 6±5 Gyr and is not well constrained (Torres, 2007). However, the rotation period of 48 days (Demory *et al.*, 2007) yields an estimated age of 2.5-3 Gyr (Barnes, 2007; Engle and Guinan, 2011). This estimation is in agreement with the lack of chromospheric activity indicated by the presence of Hα in absorption spectra.

GJ 436 has been detected by the ROSAT All-Sky Survey, which revealed an X-ray luminosity of log $L_X$ = 27.13 erg s$^{-1}$. Recent observations by the XMM-Newton spacecraft yielded a smaller value of only ~25.96 erg s$^{-1}$ (Sanz-Forcada *et al.*, 2011). We adopt the latter result because of the longer exposure time and better S/N of the XMM-Newton observations compared to ROSAT data. Using coronal models, Sanz-Forcada *et al*. (2011) extrapolated the stellar emission in the total XUV range (5 - 920Å) and found log $L_{XUV}$=26.92 erg s$^{-1}$. Adopting this value and scaling it to the HZ center at 0.24 AU, we estimate an XUV flux of about 5.14 erg cm$^{-2}$ s$^{-1}$, comparable to the present solar XUV flux at 1 AU of 4.64 erg cm$^{-2}$ s$^{-1}$ (Ribas *et al.*, 2005).

In the past, the stellar XUV flux, which was emitted from GJ 436 was certainly higher. The high-energy emission of young M dwarfs is saturated at log($L_X/L_{bol}$)~ -3 (e.g. Scalo *et al*., 2007), hence the maximum possible log($L_X$) for GJ 436 is ~29 erg s$^{-1}$. Assuming that the XUV emission of young stars occurs predominantly in X-rays because of their hotter coronae, the maximum XUV flux at the Earth-equivalent orbit around GJ 436 was roughly 700 erg cm$^{-2}$ s$^{-1}$, or ~150 $F_{XUV,now}$.

Lyman-alpha emission of GJ 436 was detected by Ehrenreich *et al.* (2011) using HST/STIS observations. They reconstructed the intrinsic stellar emission by correcting for interstellar medium (ISM) absorption, leading to an apparent flux observed at Earth of ~2.7±0.7 × 10$^{-13}$ erg cm$^{-2}$ s$^{-1}$. Scaled to the HZ center, this yields ~20.51 erg cm$^{-2}$ s$^{-1}$, a factor of ~3.3 larger than the present solar value at 1 AU of 6.19 erg cm$^{-2}$ s$^{-1}$ (Ribas *et al*., 2005). Recently, France *et al.* (2013) reconstructed the intrinsic Lyman-alpha flux of GJ 436 by using a different approach. They obtained a value of 3.5 × 10$^{-13}$ erg cm$^{-2}$ s$^{-1}$ with an uncertainty of about 15 − 30 %, which is slightly higher than the value given by Ehrenreich *et al.* (2011), but is consistent with the errors. The relevant stellar parameters and the measured radiation properties used in the model simulations for various assumed XUV flux values of a Gliese 436-type M dwarf are shown in Table 2.

The photoionization rates corresponding to the four XUV flux enhancement factors were scaled from the average present solar value at Earth (1.1 × 10$^{-7}$ s$^{-1}$; Hodges, 1994, Bzowski, 2008) to the corresponding enhancement factors. The photoionization rate is usually calculated as the product of ionization cross-section and spectral flux integrated over all wavelengths below the ionization threshold. Since full M star XUV spectra for different activity levels are currently not available, we assume that the present spectral energy distribution is equal to the solar one and scale it up by the constant factors given in Table 2.

The UV absorption rates correspond to the product of the photon flux at the center of the Lyman-alpha line and the total absorption cross-section ($5.47 \times 10^{-15}$ cm$^2$ Å; e.g. Quémerais, 2006). Adopting the reconstructed intrinsic line profile of Ehrenreich *et al.* (2011), the present value of the photon flux at the HZ center of GJ 436 is estimated to be ~$6.9 \times 10^{-3}$ s$^{-1}$.. To scale the Lyman-alpha flux and, hence, the absorption rate to higher XUV flux emission levels, we use the scaling between X-rays and Lyman-α from Ehrenreich *et al.* (2011). This assumes that $L_X/L_{bol} \approx L_{XUV}/L_{bol}$ and that the central Lyman-α flux scales approximately as the integrated line emission. The resulting absorption rates are also shown in Table 2.

### 3.2 Expected stellar wind plasma properties

Depending on the mass, size and resulting luminosity of the host star, the corresponding scaled HZ location in the case of GJ 436 is at 0.24 AU. This is a much closer orbital distance compared to that of the Earth (1 AU). As pointed out by Khodachenko *et al.* (2007), at orbital locations < 0.5 AU the flow of dense plasma related to stellar winds and coronal mass ejections (CMEs), energetic particle fluxes and XUV radiation cannot be neglected. Depending on the stand-off distance of the planetary obstacle which forms when the stellar plasma flow is deflected around the planet, previous test particle model results of Lammer *et al.* (2007) indicate that $CO_2$-rich Earth-like exoplanets having no or only weak magnetic moments may lose from tens to hundreds of bars of atmospheric pressure, or even their whole atmospheres, due to the CME induced O$^+$ ion pick-up at orbital distances $\leq 0.2$ AU.

Due to the uncertainties on the mass loss and related plasma outflow from M dwarfs, we assume in our study, as in Khodachenko *et al.* (2007) and Lammer *et al.* (2007), the plasma environment close to the stars obtained from solar observations. There exist a limited number of measurements of the mass loss rates for M stars (Wood *et al.*, 2005, Ehrenreich *et al.*, 2011), which in principle may be used for estimation of specific stellar wind density and velocity. For these stars the mass loss rates are comparable or less than the mass loss rate of the Sun. At the same time, for the Sun we have a much better knowledge of the distribution and evolution of active solar regions and of the mass outflow in the form of solar wind and CMEs, which we use in the present study. Note that stellar wind density and/or velocity lower than the solar (assumed) values would reduce the portion of the produced and picked up ions. Such a reduction would not change our main conclusion that the ion pick-up loss makes up only several percent of the thermal loss, and the further results can be considered as an upper limit for ion pick-up near an M dwarf like Gliese 436.

For the accurate study of stellar-planetary interactions one needs a reliable model which can simulate the propagation and evolution of the stellar wind plasma. For modeling the propagation and evolution of the stellar wind we use the Versatile Advection Code (VAC) (Toth, 1996). This model is able to simulate spatial and temporal evolution of the solar/stellar wind, as well as CMEs, from orbital distances $\geq 0.14$ AU. It/The model includes a self-consistent Parker-type co-rotating magnetic field, and is based on the solution of the set of the ideal (non-resistive) non-relativistic magnetohydrodynamic equations (Toth, 1996). We use a spherical uniform computation domain, which occupies a radial distance region between $0.14 < R < 1$ AU. The stellar wind in the model flows through the inner boundary of the computation domain at a semi-major axis location $d = 0.14$ AU and propagates out through the outer radial boundary at $d = 1$ AU. A self-consistent expanding stellar wind plasma flow under the conditions of a frozen-in, co-rotating, Parker-type, spiral magnetic field is numerically simulated (Odstrčil *et al.*, 1999; Rucker *et al.*, 2008).

The typical parameters of the expected flow of the stellar wind plasma at $d = 0.14$ AU are imposed along the inner radial boundary (Odstrčil *et al.*, 1999) with an initial proton concentration $n_0 = 500$ cm$^{-3}$, stellar wind proton temperature, $T_0 = 500$ kK, and a radial stellar wind velocity $v_{r0} = 300$ km s$^{-1}$. For investigating the propagation of CMEs, in a second step the simulation of a CME cloud with an initial proton density $n_0 = 1000$ cm$^{-3}$, proton temperature, $T_0 = 1700$ kK, and a radial stellar wind velocity $v_{r0} = 600$ km s$^{-1}$ is imposed at the inner boundary of the computation domain as a time-dependent injection of hot and dense plasma into the ambient stellar wind (Odstrčil *et al.*, 1999; Odstrčil *et al.*, 2004).

Fig. 3 shows radial profiles of the maximum values of plasma parameters ($n$, $v_r$, $T$) during a solar analogue CME event (dashed lines, Case II) for a typical M-star with mass $M_s \sim 0.45 M_{Sun}$ and a rotation period 2.5 days, and those for the stellar wind itself (solid lines Case I). Therefore, in the stellar wind plasma interaction modeling with the upper atmosphere, described below, two types of stellar plasma parameters representing the lower (Case I) and upper limit (Case II) conditions at the HZ location of our test planets at 0.24 AU were used as inputs:

- Case I: $n_{SW} = 250$ cm$^{-3}$, $v_{SW} = 330$ km/s, $T_{SW} = 10^6$ K,

- Case II: $n_{SW} = 700$ cm$^{-3}$, $v_{SW} = 550$ km/s, $T_{SW} = 2 \times 10^6$ K.

Here $n_{sw}$ corresponds to stellar wind concentration, $v_{sw}$ denotes its bulk velocity and $T_{sw}$ stands for the stellar wind temperature. These values cover the predicted range for the expected stellar wind plasma parameters of GJ 436 at the HZ of the system. The upper limit (Case II) can also be considered as a proxy for modeling of frequent CME events in the system when the next CME event hits the planet immediately after the previous one. We can estimate the recovery time of the planetary atmosphere after the CME event on a very simple way. Since the test planets considered in the article are not magnetized, we assume that the recovery speed coincides with the sound speed (or the speed of slow magnetoacoustic wave), so that $\tau_{CME} = R_t / c_s = R_t \left( \dfrac{\gamma k_B T_{exo}}{m_H} \right)^{-1/2}$ where $R_t$ is the obstacle width, $\gamma = 5/3$ is the ratio of the specific heats, $k_B$ is the Boltzmann constant, $m_H$ is the mass of a hydrogen atom and $T_{exo}$ is the exobase temperature. Taking the parameters from Tables 3 and 4 and depending on the characteristic size of the obstacle and the temperature, this time varies between approximately 7 and 14 hours for the Earth-type planet and 64 and 80 hours for the "super-Earth".

The next section is dedicated to the discussion of the obtained results and presents also the ion production rates for the hydrogen-rich Earth-like planet and the "super-Earth".

## 4. HYDROGEN EXOSPHERES, ENA PRODUCTION AND ION ESCAPE

In the following section we study the modification of the extended hydrogen exosphere due to the interaction of it with the stellar XUV and plasma flux. Because our hydrogen-rich test-planets orbit within a typical M-star HZ at ~0.24 AU they will be affected by tides. As it was discussed in Khodachenko *et al.* (2007) tides arise because of the finite extension of the planetary body in the inhomogeneous gravitational field of its host star so that the continuous action of the tides will reduce the planetary rotation rate or may result in a synchronous rotation (Grießmeier *et al.*, 2005).

The intrinsic magnetic field of a terrestrial planet is an essential factor for planetary protection from the ion pick-up atmosphere loss process. As shown in Khodachenko *et al.* (2007) and Lammer *et al*. (2007), because of the close orbital location within M-star HZs, the magnetosphere field of a terrestrial planet is more compressed due to the denser stellar wind impact compared with that of the solar wind at present-day Earth at 1 AU. In view of this fact the radial distance of the extended exobase of an XUV heated hydrogen rich upper atmosphere will most likely coincide with the planetary magnetopause or ionopause distance (e.g., Lammer *et al*., 2009a; Lammer *et al*., 2012, Lammer, 2013).

*4.1 Input parameters and modeling results*

As it was briefly discussed in Sect. 3.2, the magnetic moments of terrestrial planets, and especially that of "super-Earths" within close orbital distances, are expected to be weak (e.g., Gaidos *et al*., 2010; Tachinami *et al*., 2011; Morad *et al*., 2011; Stamenkovic´ et al., 2011Stamenkovi´c et al., 2012). This would lead to magnetoshperes or ionospheres which are compressed towards the planet's expanded non-hydrostatic upper atmosphere by the strong dynamic ram pressure of the dense stellar wind plasma. On the other hand if the exobase level expands beyond several planetary radii one can also expect that an Earth-type magnetosphere will not protect the exosphere (Fig. 5: Lammer *et al*., 2007; Fig. 6 right panel: Lammer *et al*., 2011a). Taking this considerations into account, the planetary obstacle most likely is located very close above the exobase level (Lammer *et al.* 2009b). Because the shape of the planetary obstacle affects the ENA production and resulting loss of planetary hydrogen ions, we study also the influence of the obstacle width on the ion production rate and the upper atmosphere erosion.

In the first case the magnetic obstacle width $R_t$ in the eq. (5) was assumed to be 1.5 times greater than the magnetopause distance. This relation and its resulting obstacle shape is close to the observed magnetosphere of the present-day Earth and may occur if the planet has an intrinsic magnetic dynamo. In the second case we choose $R_s = R_t$ so that it resembles more a Venus-type planetary obstacle which may correspond to a planet with no magnetic field, or only a very weak dynamo. By decreasing of the planetary obstacle width the ion production will significantly increase, so that a terrestrial planet with a Venus-type obstacle, where $R_s \approx R_t$, may lose larger amounts of atmospheric gas in ionized form (see Sect. 4.2). In the present study we do not specify the magnetic moment of a planet, but only chose a magnetospheric obstacle.

For illustrating the differences between the stellar wind plasma interaction with a hydrogen-rich terrestrial planet which is exposed to both a weak XUV flux and an extremely strong one, we irradiate the two test-planets with the XUV flux of the present Sun (1 XUV) and with a 100 times higher XUV flux (100 XUV). Moreover, we investigate for these two XUV flux cases several stellar wind and obstacle scenarios. The main input parameters at the inner boundary of our simulation domain are shown for 4 selected cases in Tables 3 and 4. As it was mentioned earlier, we apply the results obtained by Erkaev *et al.* (2012) to the upper atmosphere input paramters of our model. Erkaev *et al.* (2012) studied the thermosphere structure of a hydrogen-dominated Earth and a "super-Earth" ($R_{\mathrm{pl}} = 2R_{\mathrm{Earth}}$; $M_{\mathrm{pl}} = 10M_{\mathrm{pl}}$) with a hydrodynamic upper atmosphere model which solves the equations of mass, momentum and energy conservation for low and high heating efficiencies, $\eta$ of 15 % and 40 %, respectively. $\eta$ defines the percentage of the incoming XUV energy which is transferred into heating of the neutral gas. Table 4 shows the same input paramters for our model for a hotter atmosphere corresponding to the upper atmosphere values of Erkaev *et al.* (2013) with $\eta = 40\%$.

As one can see from Tables 3 and 4, the exobase distance which is chosen as our inner model boundary and the temperature are increasing for higher XUV fluxes. This behavior can be expected by taking into account a more intensive radiation flux from the parent star. Enhanced heating flux makes the scale height of the atmosphere increase, which in its turn moves the exobase to higher location. One can also see that an increase of the heating efficiency leads to an increase of the planet's obstacle stand-off (which for the studied non-magnetized planets practically coincides with the exobase) distances as well. Because of the expansion of the upper atmosphere under more extreme heating conditions (40%) the exobase densities are smaller in comparison to the 15% cases.

Figs. 4 and 5 show the modeled hydrogen exospheres and the related stellar plasma interaction under various conditions around the test planets. All the figures show cross-sections of a 3D cloud in the $x_1, x_2$ plane similar to those in Fig. 2. Fig. 4 illustrates the appearance of the extended hydrogen coronae around the Earth-type planet while Fig. 5 corresponds to the "super-Earth". In all the cases the wider Earth-type planetary obstacle is assumed, except for Fig. 4d where a smaller Venus-like obstacle is adopted.

The white area around the planets (which are shown as black dots) represents the inner atmosphere (non-hydrostatic thermosphere), which is not considered in the present study. Fig.4a and Fig.4b illustrate the influence of the XUV flux on the cloud formation. All simulation parameters for these two pictures are the same except for the XUV flux, which is chosen to be equal to and 50 times higher than the XUV flux of the present Sun. As one can see, the higher XUV flux leads to a more efficient expansion of the upper atmosphere so that charge exchange can be more intensive in the surrounding hydrogen corona. The effect of the planetary obstacle can be seen in a comparison of Fig. 4c, for an Earth-type planetary obstacle shape, with Fig.4d, for a Venus-type obstacle. In these two cases the upper atmosphere is exposed to the XUV flux which is 50 times higher compared to that of today's Sun.

Figs. 5a and 5b illustrate the importance of the heating efficiency. We show two model runs for the "super-Earth" for a comparison. Only the upper atmosphere parameters corresponding to the heating efficiencies were changed. A higher heating efficiency results in additional expansion of the upper atmosphere and in increasing production of ENAs in the vicinity of the planet (blue and red dots). Figs. 5c and 5d show the effect of the stellar wind velocity and density on the hydrogen coronae formation.

Both figures, 5c and 5d, correspond to the most extreme XUV case, which is 100 times higher than that of today's Sun.

As expected, the extreme stellar conditions result in a denser and faster stellar plasma flow, higher XUV fluxes and more intense heating of the upper atmosphere, as well as a decrease of the planetary obstacle width, which all lead to more intensive interaction processes. The ENA part of the hydrogen corona (blue and red dots) becomes more visible, meaning an increase of the atmospheric erosion processes. One can see that a huge amount of exospheric hydrogen atoms is ionized or underwent charge exchange reactions in both cases, but stronger stellar wind significantly increases the number of ENAs in the vicinity of the planet as suggested by Chassefière, (1996).

Chassefière (1996) studied the hydrodynamic outflow and escape of hydrogen atoms from a hydrogen-dominated expanded thermosphere from early Venus. From this study it was estimated that the huge ENA cloud, which is generated via charge exchange due to the interaction between an extended exosphere and the surrounding solar wind plasma of the young Sun, may contribute to about 75 % of the energy inside the thermosphere which is used for escape of the outward flowing H atoms (see Fig. 3 in Chassefière, 1996). The stellar EUV flux is deposited mainly in the lower thermosphere (Erkaev *et al*., 2013), while the ENA flux directed toward the planet should be deposited at an atmospheric layer below the exobase. It can contribute to thermospheric heating and may as a consequence modify the upper atmosphere structure which could result in an enhancement of the thermal escape rate.

Our results related to the efficient production of ENAs around the planetary obstacle support the hypothesis of Chassefière (1996) that ENAs may contribute to upper atmosphere heating. A study which investigates the possible heating contribution of ENAs additionally to the stellar XUV flux is beyond the scope of this particular work, but is in progress for a follow up study during the near future. We note also that ENA clouds near the terrestrial exoplanets within orbits around M dwarfs might be observable in the stellar Lyman-alpha line by the Hubble Space Telescope and in higher resolution beyond the geocorona in the near future by the World Space Observatory-UV (Shustov *et al.*, 2009; Lammer *et al*., 2011b).

In the next section we estimate how many of the produced planetary ions in the hydrogen coronae are picked up by the stellar wind plasma, and hence are lost from the planet.

### 4.2 Stellar wind induced atmospheric erosion of planetary hydrogen ions

As discussed above, interaction processes between the stellar wind and the upper atmosphere together with the photoionization by stellar photons lead in the case of hydrogen-rich atmospheres to the production of atmospheric $H^+$ ions, see eq. (1). After ionization, the ions can be picked up by the stellar wind plasma and swept away from the planet. Because we are interested in the efficiency of the atmospheric ion escape, we estimate the average ion production rates under various conditions. We assume as discussed below that in the considered cases the production of planetary ions and the escape rate are most likely of the same order.

Ions produced near and above the planet's obstacle can be lost because of the ion pick-up process. Since these particles are not neutral anymore, they can follow the magnetic field lines in the stellar wind plasma and can be swept away from the planet's gravity field. We consider the $H^+$ ions produced above the planetary obstacle, where the collisions between atmospheric particles can be neglected. It is assumed that the ions may be lost if the gyro radius

$$r_g = \frac{m_i v_i}{qB},$$
(6)

is small enough in comparison to the planetary radius (e.g. if the magnetic field in the vicinity of a planet is strong enough to change the trajectory of the ions significantly). Here $m_i$ is the mass of the ion, $v_i$ is the velocity of the "cold" planetary ion assumed to be ~7 km s$^{-1}$, $q$ is the ion charge and $B$ is the magnetic field near the planetary obstacle. In the case of a pure hydrogen upper atmosphere the ion mass and charge coincide with the mass and charge of a proton. The velocity of ~7 km s$^{-1}$ is chosen as being slightly faster than the mean thermal velocity of a hydrogen atom for a temperature of about 2000 K. The magnetic field at the distance ~0.24 AU from an M dwarf can be roughly estimated if one assumes a dipole character of the stellar field with the initial global value in the range of ~2 - 3 kG or ~0.2 – 0.3 T (Phan-Bao *et al.*, 2009, Reiners, 2012). By assuming an average magnetic field on the star of 2 kG, the magnetic field at 0.24 AU is $\approx 1.4 \times 10^{-3}$ G, which yields for an ion velocity of 7 km s$^{-1}$ an ion gyro radius of ~525 m which is several orders of magnitude smaller compared to the radii of the studied planets. In such a case it is justified to assume that most of the produced $H^+$ ions will be swept away from the planets by the stellar wind. If we assume that the ions are accelerated by the stellar wind electric field to the velocity of the stellar wind (Case I: 330 km/s, Case II: 550 km/s), the ion gyro radius increases proportionally to ~25 – 40 km in two extreme cases. Here we assume the filling factor *f=1* (see Phan-Bao *et al.*, 2009), i.e. the maximal possible field strength.

Since the magnetic field of ~2 - 3 kG is typical for young and active M dwarfs, it could be convenient to determine the gyro radii for a weaker field of ~50 G for an older star with the age of several Gyr (Phan-Bao *et al.*, 2009). The magnetic field at 0.24 AU is approximately $3.57 \times 10^{-5}$ G. A decrease of the magnetic field causes a proportional increase of the gyro radius (21 km for an ion velocity of 7 km s$^{-1}$, $10^3$ km and $1.6 \times 10^3$ km for 330 km/s and 550 km/s respectively). But even the highest value of ~1600 km is several times less if compared to the planetary radius, and more than an order of magnitude lower compared to the exobase radius where most ENAs are produced. These estimates support our assumption that the majority of the exospheric ions are lost from the planet and that the ion production rate is balanced by the escape rate at least during a significant part of the stellar life time.

Estimates of the ion production rates and corresponding escape rates described in the previous sections are summarized in the Tables 5 and 6 for planetary obstacles which have an Earth-like magnetosphere shape. Table 5 presents the results obtained for a hydrogen-rich Earth-like planet for two stellar wind conditions and a heating efficiency $\eta$=15% and of a higher heating efficiency $\eta = 40\%$, while Table 6 summarize the similar scenarios for the "super-Earth". Atmospheric loss rates $L_{\text{ion}}$ are given in units of particles per second. As one can see from Tables 5 and 6, in the most cases the ion production and loss rate increases with the increasing XUV flux. Loss rates for faster stellar wind also exceed the corresponding values for the wind with lower velocity which is not surprising. Since the ion production rates depend not only on the assumed heating efficiency and the XUV flux, but also on the exobase density, the values for the lower-density case corresponding to a heating efficiency of 40% (Erkaev *et al.*, 2013) are slightly lower in comparison to the 15% case shown in Table 6.

This is also the reason why the values for the cases where the stellar XUV flux is about 100 times higher than that of the present Sun are not dramatically higher (or even slightly smaller) that the ion production rates for the lower XUV fluxes, although the intensity of the interaction increases. Under the intensity of the interaction in this case we mean the ratio of produced ions to the mean exospheric density. This value increases monotonically for higher XUV fluxes in all cases considered in the present study. The reason for that is related to the corresponding exobase density that is lower because of expansion in the case of a hydrogen-rich upper atmosphere which is exposed to high XUV fluxes such as in the 100 XUV case (Erkaev *et al.*, 2013).

To investigate the influence of the planetary obstacle shape on the $H^+$ escape rate $L_{ion}$, we performed an analogous set of simulations by assuming a Venusian type planetary obstacle ( $R_s \approx R_t$ ). Tables 7 and 8 present the calculated ion production rates and estimated ion escape rates for the two test planets under the same conditions as described in Tables 5 and 6 except for the shape of the planetary obstacle. As one may see from the comparison of Tables 5 − 6 and Tables 7 − 8, reducing the width of the obstacle by 1.5 times leads to an increase in the ion production rate (i.e. the escape in general) of ≈30%. Other factors which lead to an increase of the ion production rate are related to the increase of the stellar wind density and velocity, heating efficiency $\eta$, and enhancement of the stellar XUV flux of the parent star. All these dependencies should be considered and expected.

For comparison, thermal loss rates change in the range from $4.0 \times 10^{29}$ (1 XUV, η=15%) to $4.3 \times 10^{31}$ (100 XUV, η=40%) for a hydrogen-rich Earth-type planet and from $1.6 \times 10^{29}$ (1 XUV, η=15%) to $5.3 \times 10^{31}$ (100 XUV, η=40%) for a hydrogen-rich "super-Earth". In all cases these rates exceed the presented in the current study. For more information, see Erkaev et al., (2013).

If we compare the modeled $H^+$ ion pick-up loss rates with that of Mars (e.g., Lammer *et al*., 2003) or Venus (Lammer *et al.*, 2006), which are in the order of ~$10^{25}$ s$^{-1}$, one can see that the pick-up loss rates for the hydrogen-dominated Earth-like planet are comparable for the 1 XUV case with a heating efficiency of 15 % and an Earth-like magnetopause shape, but would be a factor $10^3$ higher in the case of 40 % heating efficiency and/or a more narrower Venus-type planetary obstacle. For the larger "super-Earth" and XUV cases with higher values than that of the present-day Sun the loss rates are up to $10^4$ - $10^5$ times higher.

*4.3 Total ion escape*

The loss rates shown in Tables 5 – 8 can be used for rough estimation of the total ion loss from the hydrogen envelopes around the studied planets. This question is important because, as shown by Erkaev *et al.* (2013), volatile rich "super-Earths" which contain IR-cooling molecules can result in lower heating efficiencies of about 15 % so that for most of their lifetime they will not be in the hydrodynamic blow-off regime. In such cases non-thermal atmospheric escape processes, like the studied $H^+$ ion pick-up process, will contribute to the loss of their hydrogen-rich protoatmospheres.

Before we investigate this mass loss one needs to know how long a typical M dwarf like Gliese 436 may keep a high level of the XUV and X-ray flux. According to Penz *et al.* (2008a) the temporal scaling law for the soft X-ray (0.6-12.4 nm or 0.1-2 keV) flux of a typical M dwarf with the mass of $\approx 0.4 M_{Sun}$ can be described by the following relation

$$L_X(t \le 0.6 Gyr) \approx 0.17 L_0 t^{-0.77}, \ L_X(t > 0.6 Gyr) \approx 0.13 L_0 t^{-1.34}, \tag{6}$$

where $L_0 = 5.6 \times 10^{28}$ erg s$^{-1}$ and $t$ is given in Gyr. The soft X-ray flux can then be scaled for the appropriate orbital distance by using the relation $F_X = L_X / 4\pi d^2$.

Fig. 6 shows, approximately, the decrease of the soft X-ray flux of a Gliese 436 analogous M dwarf during the first 1.3 Gyr of its life time. Since at present time the scaling law for the XUV radiation of M dwarfs is not yet well constrained, we use this soft X-ray scaling law instead of the XUV scaling law in our estimation for the total $H^+$ escape. Assuming this temporal scaling law for the star and taking the ion production rates from Tables 5 – 8, we can estimate the mass loss from the hydrogen-rich Earth-like planet and the "super-Earth" in the HZ of a Gliese 436 type M star after 4.5 Gyr (see Tables 9 and 10). For comparison the thermal escape rates are shown by using the values from Table 2 in Erkaev *et al.* (2013).

The total ion loss is given in Earth ocean equivalent amounts of hydrogen (1EO$_H$ =1.5×10$^{23}$g). It should also be mentioned that we do not consider escape during the first ~100 Myr, the time of extreme stellar activity of an M dwarf, so that our estimates cover the time span ~0.1 – 4.5 Gyr. As discussed in Erkaev *et al.* (2013) during this early extreme period the temperature in the lower thermosphere may be >> 250 K, due to magma oceans and frequent impacts. Hot lower atmosphere will enhance the atmospheric thermal escape during this early period. A follow up study which will investigate the earliest extreme evolutionary periods is in progress.

In the present work we do not take into account the gradual decrease of the amount of gas in the planetary atmosphere due to escape processes, as we assume that the hydrogen reservoir contains much more gas in comparison to the amount lost. As discussed in Sect. 1, such scenarios can be considered as real because most of the recently discovered "super-Earths" may have huge hydrogen envelopes which contain a few % of their whole mass (e.g., Lammer, 2013).

If we compare the estimated atmospheric escape rates obtained for the thermal and ion pick-up processes for both test-planets in the HZ, it is clear that the thermal escape rate substantially exceeds the pick-up rate under the studied conditions. This is also the case when we consider the most extreme XUV and stellar plasma conditions, including a narrow planetary obstacle. A fraction of the evaporating neutral exosphere will be ionized so that ion pick-up will contribute to the total atmospheric escape rate, but as long as the upper atmosphere is in blow-off thermal escape will be the most important.

In our study the escape rate of ionized hydrogen atoms can vary for an Earth-type planet from ~0.6 $EO_H$ under moderate conditions (15% heating efficiency, moderate stellar wind, Earth-type planetary obstacle) up to ~2.5 $EO_H$ for extreme environments (40% heating efficiency, denser and faster wind and a narrow Venusian obstacle type).

The stronger gravity of the more massive "super-Earth" keeps the atmosphere closer to the planet's surface, which slows down the charge exchange and photoionization processes. In this case the escape changes from ~0.4 $EO_H$ to ~2.14 $EO_H$ depending on the environment but remains smaller compared to escape from an Earth-type planet (ranging from ~0.6 $EO_H$ to 2.5 $EO_H$, see above). In all studied cases the non-thermal $H^+$ pick-up rate is several times smaller in comparison with the thermal one, but makes up a significant fraction of the whole loss processes.

The results of our study, together with those of Erkaev *et al.* (2013), indicate that terrestrial exoplanets ranging in mass and size from Earth- to "super-Earths" may experience difficulties in losing dense hydrogen envelopes if they have H-dominated protoatmosphere remnants with >9 $EO_H$ (Earth: $\eta = 15\%$) and >19 $EO_H$ (Earth: $\eta = 40\%$).

For "super-Earth" these amounts are >3.5 $EO_H$ ($\eta$ = 15%) and >10 ($\eta$ = 40%). Dense hydrogen envelopes may be removed more easily if a particular exoplanet is located closer to its parent star such as Corot-7b or Kepler-10b (e.g., Leitzinger *et al.*, 2011) which orbit their parent stars at ~ 0.017 AU, but not inside the HZ. We note also that due to the HZ location at greater distances from the parent stars, the stellar wind erosion of hydrogen-envelopes will be less efficient for planets inside the HZs of K, G and F-type stars.

## 5. CONCLUSIONS

In this study we investigated the non-thermal ion pick-up escape process from hydrogen-rich, non-hydrostatic upper atmospheres of an Earth-like planet, and a hydrogen-rich "super-Earth" which is twice as large as Earth and ten times more massive. Both planets are supposed to be located inside the HZ of a typical M dwarf star with stellar properties similar as GJ 436. We showed that in the case of an M dwarf the produced planetary $H^+$ ions have a high probability to be picked up from the extended hydrogen coronae by the stellar wind plasma flow so that the ion production rate and the ion escape rate are perhaps of the same order. We exposed the two test-planets to various XUV fluxes from 1 to 100 times that of the present Sun and found that in all studied cases the ionization of exospheric neutral hydrogen atoms by charge-exchange and photoionization contributes to the total atmospheric escape of the upper atmosphere, but do not prevail over the thermal escape. The total non-thermal atmospheric escape by ion pick-up from possible dense hydrogen envelopes during the life time of the studied planets is < 3 $EO_H$. Our results indicate that if a rocky exoplanet did not lose the majority of its nebula captured hydrogen gas envelope, or degassed a huge amount of hydrogen-rich volatiles by thermal blow-off during the first hundred Myr after the planet's origin, it is questionable if the stellar wind can erode a remaining dense hydrogen envelope non-thermally. The thermal escape is higher during the planet's history, but is probably unable to remove such dense hydrogen envelopes as well. The situation may change dramatically for exoplanets which are located closer to their parent stars. Depending on the nebula life time, the formation process, planetary mass, stellar activity and plasma properties in the vicinity of the planet, the initial amount of hydrogen, thermal and non-thermal atmospheric escape processes will determine if a planet becomes a world with an Earth-type atmosphere (and no hydrogen envelope) or remains a sub-Neptune type body.

ACKNOWLEDGEMENTS

M. Güdel, K. G. Kislyakova, M. L. Khodachenko, and H. Lammer acknowledge the support by the FWF NFN project S116 "Pathways to Habitability: From Disks to Active Stars, Planets and Life", and the related FWF NFN subprojects, S116 604-N16 "Radiation & Wind Evolution from T Tauri Phase to ZAMS and Beyond", S116 606-N16 "Magnetospheric Electrodynamics of Exoplanets", S116607-N16 "Particle/Radiative Interactions with Upper Atmospheres of Planetary Bodies Under Extreme Stellar Conditions". K. G. Kislyakova, Yu. N. Kulikov, H. Lammer, and P. Odert thank also the Helmholtz Alliance project "Planetary Evolution and Life". P. Odert and M. Leitzinger acknowledges support from the FWF project P22950-N16. The authors also acknowledge support from the EU FP7 project IMPEx (No.262863) and the EUROPLANET-RI projects, JRA3/EMDAF and the Na2 science WG5. The authors thank the International Space Science Institute (ISSI) in Bern, and the ISSI team "Characterizing stellar- and exoplanetary environments". Finally, N. V. Erkaev acknowledges support by the RFBR grant No 12-05-00152-a. This research was conducted using resources provided by the Swedish National Infrastructure for Computing (SNIC) at the High Performance Computing Center North (HPC2N), Umeå University, Sweden. The software used in this work was in part developed by the DOE-supported ASC / Alliance Center for Astrophysical Thermonuclear Flashes at the University of Chicago. Finally, the authors thank the referees for very useful suggestions and recommendations which helped to improve the work.

**REFERENCES**

Barnes, S. A. (2007) Ages for Illustrative Field Stars Using Gyrochronology: Viability, Limitations, and Errors, *ApJ*, 669, 1167-1189.

Ben-Jaffel, L., Sona Hosseini, S. (2010) On the existence of energetic atoms in the upper atmosphere of exoplanet HD 209458b, *ApJ*, 709, 1284–1296.

Bonfils, X., Delfosse, X., Udry, S., Forveille, T., Mayor, M., Perrier, C., Bouchy, F., Gillon, M., Lovis, C., Pepe, F., Queloz, D., Santos, N. C., Ségransan, D., Bertaux, J.-L. (2011) The HARPS search for southern extra-solar planets XXXI. The M dwarf sample, *A&A,* submitted, 2011arXiv1111.5019B.

Butler, R. P., Vogt, S. S., Marcy, G. W., Fischer, D. A., Wright, J. T., Henry, G. W., Laughlin, G., Lissauer, J. J. (2004) A Neptune-Mass Planet Orbiting the Nearby M Dwarf GJ 436, *ApJ*, 617, 580-588.

Bzowski, M. (2008) Survival probability and energy modification of hydrogen energetic neutral atoms on their way from the termination shock to Earth orbit, *A&A*, 488, 1057-1068.

Chandrasekhar, S. (1963) The equilibrium and the stability of the Roche ellipsoids, *ApJ,* 138, 1182-1213.

Chassefière, E. (1996) Hydrodynamic escape of hydrogen from a hot water-rich atmosphere: the case of Venus, *J. Geophys. Res.,* 101, E11, 26,039-26,056.

Charbonneau, D., Berta, Z. K., Irwin, J., Burke, C. J., Nutzman, P., Buchhave, L. A., Lovis, C., Bonfils, X., Latham, D.W., Udry, S., Murray-Clay, R. A., Holman, M. J., Falco, E. E.,Winn, J. N., Queloz, D., Pepe, F., Mayor, M., Delfose, X., Forveille, T. (2009) A super-Earth transiting a nearby low-mass star, *Nature*, 462, 891–894.

Chaufray, J. Y., Modolo, R., Leblanc, F., Chanteur, G., Johnson, R. E., Luhmann, J. G. (2007) Mars solar wind interaction: formation of the Martian corona and atmospheric loss to space. *J. Geophys. Res*., 112, Issue E9, CiteID E09009.

Demory, B.-O., Gillon, M., Barman, T., Bonfils, X., Mayor, M., Mazeh, T., Queloz, D., Udry, S., Bouchy, F., Delfosse, X., Forveille, T., Mallmann, F., Pepe, F., Perrier, C.

(2007) Characterization of the hot Neptune GJ 436 b with Spitzer and ground-based observations, *A&A*, 475, 1125-1129.

Dressing, C. D., Charbonneau, D., (2013) The occurrence rate of small planets around small stars. *ApJ*, in press, eprint arXiv: 2013arXiv1302.1647D.

Ehrenreich, D., Lecavelier des Etangs, A., Delfosse, X. (2011) HST/STIS Lyman-α observations of the quiet M dwarf GJ 436. Predictions for the exospheric transit signature of the hot Neptune GJ 436b, *A&A*, 529, A80.

Ehrenreich, D., Bourrier, V., Bonfils, X., Lecavelier des Etangs, A., Hébrard, G., Sing, D. K., Wheatley, P. J., Vidal-Madjar, A., Delfosse, X., Udry, S., Forveille, T., Moutou, C. (2012) Hint of a transiting extended atmosphere on 55 Cancri b, *A&A*, 547, A18.

Ekenbäck, A., Holmström, M., Wurz, P., Grießmeier, J.-M., Lammer, H., Selsis, F., Penz, T. (2010) Energetic neutral atoms around HD 209458b: Estimations of magnetospheric properties, *ApJ*, 709, 670–679.

Elkins-Tanton, L. T. (2011) Formation of water ocean on rocky planets, *Astrophys. Space Sci.*, 332, 359–364.

Endl, M., Robertson, P., Cochran, W. D., MacQueen, P. J., Brugamyer, E. J., Caldwell, C., Wittenmyer, R. A., Barnes, S. I., Gullikson, K. (2012) Revisiting $\rho^1$ Cancri e: A new mass determination of the transiting super-Earth, *ApJ*, 759, 19.

Engle, S. G., Guinan, E. F. (2011) Red Dwarf Stars: Ages, Rotation, Magnetic Dynamo Activity and the Habitability of Hosted Planets, *ASPC*, 451, 285-296.

Erkaev, N. V., Penz, T., Lammer, H., Lichtenegger, H. I. M.,Wurz, P., Biernat, H. K., Griessmeier, J.-M., Weiss, W. W. (2005) Plasma and magnetic field parameters in the vicinity of short periodic giant  exoplanets, *ApJ,* 157, 396–401.

Erkaev, N. V., Lammer, H., Odert, P., Kulikov, Yu. N., Kislyakova, K. G., Khodachenko, M. L., Güdel, M., Hanslmeier, A., Biernat, H. K. (2013) XUV exposed non-hydrostatic hydrogen-rich upper atmospheres of Earth- and "super-Earth"-size planets. Part I: Atmospheric expansion and thermal escape, *Astrobiology*, submitted.

France, K., Froning, C. S., Linsky, J. L., Roberge, A., Stocke, J. T., Tian, F., Bushinsky, R., Desert, J.-M., Mauas, P., Vieytes, M., Walcowicz, L. M. (2013) The ultraviolet radiation environment around M dwarf exoplanet host stars, *ApJ*, 763, 14.

Fox, J. L., Hać, A. B. (2009) Photochemical escape of oxygen from Mars: a comparison of the exobase approximation to a Monte Carlo method. Icarus 204, 527–544.

Fryxell, B., Olson, K., Picker, R., Timmes, F.X., Zingale, M., Lamb, D.Q., MacNeice, P., Rosner, R., Truran, J.W., Tufo, H. (2000) FLASH: an adaptive MESH hydrodynamic code for modeling astrophysical thermonuclear flashes, *ApJ Supplement Series*, 131, 273–334.

Fuselier, S.A., Funsten, H.O., Heirtzler, D., Janzen, P., Kucharek, H., McComas, D.J., Möbius, E., Moore, T.E., Petrinec, S.M., Reisenfeld, D.B., Schwadron, N.A., Trattner, K.J., Wurz, P. (2010) Energetic neutral atoms from the Earth's subsolar magnetopause. *Geophys. Res. Lett.,* 77, L13101, 1-5.

Futaana, Y., Barabash, S., Grigoriev, A., Holmström, M., Kallio, E., Brandt, P.C.:Son, Gunell, H., Brinkfeldt, K., Lundin, R., Andersson, H., Yamauchi, M., McKenna-Lawler, S., Winningham, J. D., Frahm, R.A., Sharber, J. R., Scherrer, J. R., Coates, A. J., Linder, D. R., Kataria, D. O., Säles, T., Riihelä, P., Schmidt, W., Koskinen, H., Kozyra, J., Luhmann, J., Roelof, E., Williams, D., Livi, S., Curtis, C. C., Hsieh, K. C., Sandel, B. R., Grande, M., Carter, M., Sauvaud, J.-A., Fedorov, A., Thocaven, J.-J., Orsini, S., Cerulli-Irelli, R., Maggi, M., Wurz, P., Bochsler, P., Galli, A., Krupp, N., Woch, J., Fränz, M.,

Asamura, K., Dierker, C. (2006) First ENA observations at Mars: Subsolar ENA jet. *Icarus*, 182, 413–423.

Gaidos, E., Conrad, C. P., Manga, M., Hernlund, J. (2010) Thermodynamic limits on magnetodynamos in rocky exoplanets. *ApJ*, 718, 598-609.

Galli, A., Wurz, P., Bochsler, P., Barabash, S., Grigoriev, A., Futaana, Y., Holmström, M., Gunell, H., Andersson, H., Lundin, R., Yamauchi, M., Brinkfeldt, K., Fränz, M., Krupp, N., Woch, J., Baumjohann, W., Lammer, H., Zhang, T. L., Asamura, K., Coates, A. J., Linder, D. R., Kataria, D. O., Curtis, C. C., Hsien, K. C., Sandel, B. R., Sauvaud, J. A., Fedorov, A., Mazelle, C., Thocaven, J. J., Grande, M., Kallio, E., Sales, T., Schmidt, W., Riihela, P., Koshkinen, H., Kozyra, J., Luhmann, J., Mc-Kenn-Lawlor, S., Orsini, S., Cerulli-Irelli, R., Mura, A., Milillo, A., Maggi, M., Roelof, E., Brandt, P., Russel, C. T., Szego, K., Winningham, D., Frahm, R., Scherrer, J., Sharber, J. R. (2008) First observation of energetic neutral atoms in the Venus environment. *Planet. Space Sci.*, 56, Issue 6, 807–811.

García Muñoz, A. (2007) Physical and chemical aeronomy of HD 209458b, *Planet. Space Sci.*, 55, 1426–1455.

Gillon, M., Pont, F., Demory, B.-O., Mallmann, F., Mayor, M., Mazeh, T., Queloz, D., Shporer, A., Udry, S., Vuissoz, C. (2007) Detection of transits of the nearby hot Neptune GJ 436 b, *A&A*, 472, L13-L16.

Grießmeier, J.-M., Stadelmann, A., Motschmann, U., Belisheva, N.K., Lammer, H., Biernat, H.K. (2005) Cosmic ray impact on extrasolar Earth-like planets in close-in Habitable Zones. *Astrobiology*, 5, Issue 5, 587-603.

Gröller, H., Shematovich,V. I., Lichtenegger, H. I. M., Lammer, H., Pfleger, M., Kulikov, Yu. N., Macher, W., Amerstorfer, U. V., Biernat, H. K. (2010) Venus' atomic hot oxygen environment, *J. Geophys. Res*., 115, E12017.

Gröller, H., Lammer, H., Lichtenegger, H.I.M., Pfleger, M., Dutuit, O., Shematovich, V.I., Kulikov, Yu.N., Biernat, H.K. (2012) Hot oxygen atoms in the Venus nightside exosphere, *Geophys. Res. Lett*., 39, L03202.

Hayashi, C., Nakazawa K., Mizuno H. (1979) Earth's melting due to the blanketing effect of the primordial dense atmosphere. *Earth Planet. Sci. Lett*., 43, 22-28.

Hodges, R., Richard, Jr., (1994) Monte Carlo simulation of the terrestrial hydrogen exosphere, *JGR*, 99, 23229.

Holmström, M., Ekenbäck, A., Selsis, F., Penz, T., Lammer, H., Wurz, P. (2008) Energetic neutral atmos as the explanation for the high-velocity hydrogen around HD 209458b, *Nature*, 451, 970-972.

Ikoma, M., Genda, H., (2006) Constraints on the mass of a habitable planet with water of nebular origin. *ApJ*, 648, 696-706.

Ikoma, M., Hory, Y. (2012) In situ accretion of hydrogen-rich atmospheres on short-period Super-Earths: implications for the Kepler-11 planets. *ApJ*, 753, Issue 1, 6.

Johnson, R. E. (1990) *Energetic Charged Particle Interactions with Atmospheres and Surfaces*, Springer, Berlin.

Kasting, J. F., Pollack, J. B. (1983) Loss of water from Venus. I. Hydrodynamic escape of hydrogen, *Icarus*, 53, 479-509.

Khodachenko, M.L., Ribas, I., Lammer, H., Grießmeier, J.-M., Leitner, M., Selsis, F., Eiroa, C., Hanslmeier, A., Biernat, H., Farrugia, C.J., Rucker, H. (2007) Coronal Mass Ejection (CME) Activity of low mass M stars as an important factor for the habitability of terrestrial exoplanets. I. CME impact on expected magnetospheres of Earth-like exoplanets in close-in habitable zones, *Astrobiology,* 7, 167-184.

Korenaga, J. (2010) On the likelihood of plate tectonics on super-Earths: Does size matter? *ApJL*, 725, L43 − L46.

Koskinen, T. T., Harris, M. J., Yelle, R. V., Lavvas, P. (2012) The escape of heavy atoms from the ionosphere of HD 209458b. I. A photochemical-dynamical model of the thermosphere, *Icarus*, accepted, http://arXiv:1210.1535.

Krestyanikova, M. A., Shematovich, V. I. (2005) Stochastic models of hot planetary and satellite coronas: a photochemical source of hot oxygen in the upper atmosphere of Mars, *Sol. Syst. Res*., 39, 22 - 32.

Krestyanikova, M.A., Shematovich, V.I. (2006) Stochastic models of hot planetary and satellite coronas: a hot oxygen corona of Mars, *Sol. Syst. Res.*, 40, 384–392.

Kuchner, M. J. (2003) Volatile-rich Earth-mass planets in the habitable zone, *ApJ*, 596, L105-L108.

Lammer, H., Lichtenegger, H.I.M., Kulikov, Yu.N., Grießmeier, J.-M., Terada, N., Erkaev, N.V., Biernat, H.K., Khodachenko, M.L., Ribas, I., Penz, T., Selsis, F. (2007) Coronal Mass Ejection (CME) activity of low mass M stars as an important factor for the habitability of terrestrial exoplanets. II. CME-induced ion pick-up of Earth-like exoplanets in close-in habitable zones, *Astrobiology*, 7, 185-207.

Lammer, H., Bredehöft, J. H., Coustenis, A., Khodachenko, M. L., Kaltenegger, L., Grasset, O., Prieur, D., Raulin, F., Ehrenfreund, P., Yamauchi, M., Wahlund, J.-E., Grießmeier, J.-M., Stangl, G., Cockell, C. S., Kulikov, Y. N., Grenfell, J. L., Rauer H. (2009a) What makes a planet habitable?, *A&A*, 17, 181–249.

Lammer, H., Lichtenegger, H. I. M., Kolb, C., Ribas, I., Guinan, E. F., Bauer, S. J. (2003) Loss of water from Mars: Implications for the oxidation of the soil, *Icarus*, 165, 9-25.

Lammer, H., Lichtenegger, H.I.M., Biernat, H. K., Erkaev, N. V., Arshukova, I. L., Kolb, C., Gunell, H., Lukyanov, A., Holmstrom, M., Barabash, S., Zhang, T. L., Baumjohann, W. (2006) Loss of hydrogen and oxygen from the upper atmosphere of Venus, *Planet. Space Sci.*, 54, 1445-1456.

Lammer, H., Odert, P., Leitzinger, M., Khodachenko, M. L., Panchenko, M., Kulikov, Yu. N., Zhang, T.L., Lichtenegger, H.I.M., Erkaev, N.V., Wuchterl, G., Penz, T., Biernat, H.K., Weingrill, J., Steller, M., Ottacher, H., Hasiba, J., Hanslmeier, A. (2009b) Determining the mass loss limit for close-in exoplanets: what can we learn from transit observations? *A&A*, 506, 399–410.

Lammer, H., Kislyakova, K. G., Odert, P., Leitzinger, M., Schwarz, R., Pilat-Lohinger, E., Kulikov, Yu. N., Khodachenko, M. L., Güdel, M., Hanslmeier, A. (2011a) Pathways to Earth-like atmospheres: extreme ultraviolet (EUV)-powered escape of hydrogen-rich protoatmospheres, *Orig.Life Evol. Biosph.*, 41, 503–522.

Lammer, H., Eybl, V., Kislyakova, K. G., Weingrill, J., Holmström, M., Khodchenko, M. L., Kulikov, Yu. N., Reiners, A., Leitzinger, M., Odert, P., Xian Grüß, M., Dorner, B., Güdel, M., Hanslmeier, A. (2011b) UV transit observations of EUV-heated expanded thermospheres of Earth-like exoplanets around M-stars: Testing atmosphere evolution scenarios, *Astrophys. Space Sci.*, 335, 39–50.

Lammer, H. (2013) *Origin and evolution of planetary atmospheres: implications for habitability*, Springer Briefs in Astronomy, Springer Verlag, Heidelberg, New York.

Leblanc, F., Johnson, R. E. (2002) Role of molecular species in pick-up ion sputtering of the Martian atmosphere, *J. Geophys. Res.*, 107, 1–6.

Lecavelier des Etangs, A., Ehrenreich, D., Vidal-Madjar, A., Ballester, G. E., Désert, J.-M., Ferlet, R., H´ebrard, G., Sing, D. K., Tchakoumegni, K.-O., Udry, S. (2010) Evaporation of the planet HD 189733b observed in H I Lyman-α, *A&A*, 514, A72.

Léger, A., Selsis, F., Sotin, C., Guillot, T., Despois, D., Mawet, D., Ollivier, M., Labèque, A., Valette, C., Brachet, F., Chazelas, B., Lammer H. (2004) A new family of planets? "Ocean-Planets", *Icarus*, 169, 499–504.

Leitzinger, M., Odert, P., Kulikov, Yu. N., Lammer, H., Wuchterl, G., Penz, T., Guarcello, M. G., Micela, G., Khodachenko, M. L., Weingrill, J., Hanslmeier, A., Biernat, H. K., Schneider, J. (2011) Could CoRoT-7b and Kepler-10b be remnants of evaporated gas or ice giants? *Planet. Space Sci.*, 59, Issue 13, 1472-1481.

Lichtenegger, H. I. M., Lammer, H., Grießmeier, J.-M., Kulikov, Yu. N., von Paris, P., Hausleitner, W., Krauss, S., Rauer, H. (2010) Aeronomical evidence for higher $CO_2$ levels during Earth's Hadean epoch, *Icarus*, 210, 1-7.

Linsky, J. L., Yang, H., France, K., Froning, C. S., Green, J. C., Stocke, J. T., Osterman, S. N. (2010) Observations of mass loss from the transiting exoplanet HD 209458b. ApJ 717, 1291–1299.

Lissauer, J. J., and the Kepler team (2011) A closely packed system of low-mass, low-density planets transiting Kepler-11, *Nature*, 470, 53–58.

Lundin, R., Lammer, H., Ribas, I. (2007) Planetary magnetic fields and solar forcing: implications for atmospheric evolution, *Space Science Reviews*, 129, 245-278.

Ma, Y.-J., Nagy, A. F. (2007) Ion escape fluxes from Mars. *Geophys. Res. Lett.*, 34, L08201.

Maness, H. L., Marcy, G. W., Ford, E. B., Hauschildt, P. H., Shreve, A. T., Basri, G. B., Butler, R. P., Vogt, S. S. (2007) The M Dwarf GJ 436 and its Neptune-Mass Planet, *PASP*, 119, 90-101.

Mizuno, H., Nakazawa, K., Hayashi, C. (1978) Instability of a gaseous envelope surrounding a planetary core and formation of giant planets. *Prog. Theor. Phys.*, 60, 699–710.

Morard, G., Bouchet, J., Valencia, D., Mazevet, S., Guyot, F. (2011) The melting curve of iron at extreme pressures: Implications for planetary cores. *High Energy Density Phys.*, 7, 141-144.

Möstl, U.V., Erkaev, N.V., Zellinger, M., Lammer, H., Gröller, H., Biernat, H.K., Korovinskiy, D. (2011) The Kelvin-Helmholtz instability at Venus: what is the unstable boundary? *Icarus*, 216, 476–484.

Murray-Clay, R. A., Chiang, E. I., Murray, N. (2009) Atmospheric escape from Hot Jupiters, *ApJ,* 693, 23-42.

Odstrčil, D., Pizzo, V.J. (1999) Three-dimensional propagation of CMEs in a structured solar wind flow: 1. CME launched within the streamer belt, *J. Geophys. Res.*, 104, A1:483-492.

Odstrčil, D., Riley, P., Zhao, X.P. (2004) Numerical simulation of the 12 May 1997 interplanetary CME event, *J. Geophys. Res.,* 109, A2, CiteID A0211.

Penz, T., Erkaev, N. V., Biernat, H. K., Lammer, H., Amerstorfer, U. V., Gunell, H., Kallio, E., Barabash, S., Orsini, S., Milillo, A., Baumjohann, W. (2004) Ion loss on Mars caused by the Kelvin-Helmholtz instability, *Planet. Space Sci.*, 52, 1157–1167.

Penz, T., Michela, G., Lammer, H. (2008a) Influence of the evolving stellar X-ray luminosity distribution on exoplanetary mass loss, *A&A*, 477, 309-314.

Penz, T., Erkaev, N. V., Kulikov, Yu. N., Langmayr, D., Lammer, H., Micela, G., Cecchi-Pestellini, C., Biernat, H. K., Selsis, F., Barge, P., Deleuil, M., Léger, A., (2008b) Mass loss from "Hot Jupiters" - Implications for CoRoT discoveries, Part II: Long time thermal atmospheric evaporation modelling, *Planet. Space Sci.*, 56, 1260–1272.

Phan-Bao, N., Lim, J., Donati, J.-F., Johns-Krull, C.M., Martin, E.L. (2009) Magnetic field topology in low-mass stars: spectropolametric observations of M dwarfs, *ApJ*, 704, 1721-1729.

Prölss, G. W. (2004) Physik des erdnahen Weltraums, *Publishing House Heidelberg New York*, pp. 529.

Quémerais, E. (2006) The Interplanetary Lyman-alpha Background, in: *The Physics of the Heliospheric Boundaries*, ISSI Scientific Report No. 5, 283-310.

Rafikov, R. R., (2006) Atmospheres of protoplanetary cores: critical mass for nucleated instability. *ApJ.*, 648, 666-682.

Reiners, A. (2012) Observations of Cool-Star Magnetic Fields, *Living Rev Solar Phys.*, 8, 1.

Ribas, I., Guinan, E. F., Güdel, M., Audard, M. (2005) Evolution of the Solar Activity over Time and Effects on Planetary Atmospheres. I. High-Energy Irradiances (1-1700Å), *ApJ*, 622, 680-694.

Rucker, H.O., Panchenko, M., Hansen, K.C., Taubenschluss, U., Boudjada, M.Y., Kurth, W.S., Dougherty, M.K., Steinberg, J.T., Zarka, P., Galopeau, P.H.M., McComas, D.J.,

Barrow, C.H. (2008) Saturn kilometric radiation as a monitor for the solar wind? *Adv. Space Res.*, 42, 40-47.

Sanz-Forcada, J., Micela, G., Ribas, I., Pollock, A. M. T., Eiroa, C., Velasco, A., Solano, E., Garca-Ã•lvarez, D. (2011) Estimation of the XUV radiation onto close planets and their evaporation, *A&A*, 532, A6.

Scalo, J., Kaltenegger, L., Segura, A., Fridlund, M., Ribas, I., Kulikov, Y. N., Grenfell, J. L., Rauer, H., Odert, P., Leitzinger, M., Selsis, NFN project F., Khodachenko, M. L., Eiroa, C., Kasting, J., Lammer, H. (2007) M stars as targets for terrestrial exoplanet searches and biosignature detection, *Astrobiology*, 7, 85 – 166.

Selsis, F., Kasting, J. F., Levrard, B., Paillet, J., Ribas, I., Delfosse, X. (2007) Habitable planets around the star Gliese 581? *A&A*, 476, 1373-1387.

Shematovich, V. I. (2010) Suprathermal hydrogen produced by the dissociation of molecular hydrogen in the extended atmosphere of exoplanet HD 209458b, *Sol. Syst. Res.*, 44, 96–103.

Shustov, B., Sachov, M., Gomez de Castro, A. I., Ana, I., Pagano, I. (2009) WSO-UV ultraviolet mission for the next decade. *Astrophys. Space Sci.*, 320, 187–190.

Stamenkovic´, V., Breuer, D., Spohn, T. (2011) Thermal and transport properties of mantle rock at high pressure: Applications to super-Earths. *Icarus*, 216, 572-596.

Stamenkovic´, V., Noack, L., Breuer, D., Spohn, T. (2012) The influence of pressure-dependent viscosity on the thermal evolution of super-Earths. *ApJ*, 748, 22.

Tachinami, C., Senshu, H., Ida, S. (2011) Thermal evolution and lifetime of intrinsic magnetic fields of super-Earths in habitable zones. *ApJ*, 726, 18.

Terada, N., Machida, S., Shinagawa, H. (2002) Global hybrid simulation of the Kelvin-Helmholtz instability at the Venus ionopause, *J. Geophys. Res*., 107, 1471–1490.

Tian, F., Toon, O. B., Pavlov, A. A., De Sterck, H. (2005a) Transonic hydrodynamic escape of hydrogen from extrasolar planetary atmospheres. *ApJ*, 621, 1049–1060.

Tian, F., Toon, O. B., Pavlov, A. A., De Sterck, H. (2005b) A hydrogen-rich early Earth atmosphere, *Science*, 308, 1014–1017.

Tian, F., Kasting J. F., Liu, H., Roble, R. G. (2008a) Hydrodynamic planetary thermosphere model: 1. The response of the Earth's thermosphere to extreme solar EUV conditions and the significance of adiabatic cooling, *J. Geophys. Res*., 113, E05008.

Tian, F., Solomon, S. C., Qian, L., Lei, J., Roble, R. G. (2008b) Hydrodynamic planetary thermosphere model: 2. Coupling of an electron transport/energy deposition model, *J. Reophys. Res.*, 113, E07005.

Torres, G. (2007) The Transiting Exoplanet Host Star GJ 436: A Test of Stellar Evolution Models in the Lower Main Sequence, and Revised Planetary Parameters, *ApJ*, 671, L65-L68.

Toth, G. (1996) A General Code for Modeling MHD Flows on Parallel Computers: Versatile Advection Code, *Astrophys. Lett. Com.*, 34, 245.

Valencia, D., O'Connell, R. J., Sasselov, D. D. (2007) Inevitability of plate tectonics on super-Earths, *ApJ*, 670, L45 - L48.

Van Heck, H. J., Tackley, P. J. (2011) Plate tectonics on super-Earths: Equally or more likely than on Earth, *Earth Planet. Sci. Lett*., 310, 25

Vidal-Madjar, A., Lecavelier des Etangs, A., Désert, J.-M., Ballester, G.E., Ferlet, R., Hebrard, G., Mayor, M (2003) An extended upper atmopshere around the extrasolar planet HD 209458b, *Nature*, 422, 143-146.

Volkov, A. N., Johnson, R. E., Tucker, O. J., Erwin, J. T. (2011) Thermally driven atmospheric escape: Transition from hydrodynamic to Jeans escape, *ApJ*, 729, art. id. L24.

Von Braun, K., Boyajian, T. S., Kane, S. R., Hebb, L., van Belle, G. T., Farrington, C. C., David R. Knutson, H. A., ten Brummelaar, T. A., López-Morales, 9.119864e+2M., McAlister, H. A., Schaefer, G., Ridgway, S., Collier Cameron, A., Goldfinger, P. J.; Turner, N. H., Sturmann, L., Sturmann, J. (2012) The GJ 436 system: Directly determined astrophysical parameters of an M dwarf and implications for the transiting hot Neptune, *ApJ*, 753, Issue 2, article id. 171.

Watson, A. J., Donahue, T. M., Walker, J. C. G. (1981) The dynamics of a rapidly escaping atmosphere - Applications to the evolution of earth and Venus, *Icarus*, 48, 150 – 166.

Wei, Y., Fraenz, M., Dubinin, E., Woch, J., Lühr, H., Wan, W., Zong, Q.-G., Zhang, T.-L., Pu, Z. Y., Fu, S. Y., Barabash, S., Lundin, R., Dandouras, I. (2012) Enhanced atmospheric oxygen outflow on Earth and Mars driven by a corotating interaction region. *J. Geophys. Res*., 117, A03208.

Wood, B. E., Müller, H.-R., Zank, G. P., Linsky, J. L., Redfield, S. (2005) New mass-loss measurements from Asrospheric Ly-alpha absorption. *ApJ*, 628, Issue 2, L143-L146.

Wuchterl, G. (1993) The critical mass for protoplanets revised - Massive envelopes through convection. *Icarus*, 106, 323–334.

Yau, A. W., André, M. (1997) Sources of ion outflow in the high latitude ionosphere, *Space Sci. Rev.*, 37, 1–25.

Yelle, R. V. (2004) Aeronomy of extra-solar giant planets at small orbital distances, *Icarus*, 170, 167– 179.

Yelle, R. (2006) Corrigendum to Aeronomy of extra-solar giant planets at small orbital distances (Icarus 170 (2004) 167–179). *Icarus*, 183, 508.

## TABLES

**Table 1.** Model inner boundary "ib" atmospheric and solar wind input parameters for present-day Earth's geocorona (Prölss, 2004).

| $R_{ib}$ [$R_{Earth}$] | $n_{Hib}$ [cm$^{-3}$] | $T_{ib}$ [K] | XUV$_{Sun}$ | $n_{sw}$ [cm$^{-3}$] | $v_{sw}$ [km s$^{-1}$] | $T_{sw}$ [K] |
|---|---|---|---|---|---|---|
| 1.078 | $7 \times 10^4$ | 900 | 1 | 8 | 400 | $1 \times 10^6$ |

**Table 2**. Default stellar parameters as well as values of constants used in the simulations.

| Stellar radius [$R_{Sun}$] | 0.45 |
|---|---|
| Stellar mass [$M_{Sun}$] | 0.45 |
| Habitable zone [AU] | $0.12 - 0.36$ AU |
| Orbital distance [AU] | 0.24 |
| Angular velocity [rad s$^{-1}$] | $1.14 \times 10^{-6}$ |
| Orbital period [days] | 64 |
| XUV flux exposure [XUV$_{Sun}$] | 1 \| 10 \| 50 \| 100 |
| UV absorption rate [s$^{-1}$] | $3.9 \times 10^{-3}$ \| $1.2 \times 10^{-2}$ \| $2.7 \times 10^{-2}$ \| $3.9 \times 10^{-2}$ |
| Photoionization rate [s$^{-1}$] | $1.1 \times 10^{-7}$ \| $1.1 \times 10^{-6}$ \| $5.5 \times 10^{-6}$ \| $1.1 \times 10^{-5}$ |

**Table 3**. Planetary input parameters taken from Erkaev *et al.* (2013) for different stellar wind properties and XUV fluxes which correspond to that of the present Sun and an XUV enhancement factor of about 100 times for a low heating efficiency $\eta = 15\%$.

| Parameter / Case | Earth-type planet, 1 XUV, Case I | Earth-type planet, 100 XUV, Case II | "super-Earth", 1 XUV, Case I | "super-Earth", 100 XUV, Case II |
|---|---|---|---|---|
| Inner boundary: [$R_{pl}$] | 7.4 | 18.8 | 3.5 | 13.0 |
| Inner boundary T [K] | 240 | 2310 | 100 | 2075 |
| Obstacle stand-off distance:[$R_{pl}$] | 7.5 | 19 | 3.6 | 13.1 |
| Obstacle width: [$R_{pl}$] | 15 | 28.5 | 5.3 | 19.6 |
| Inner boundary density: [cm$^{-3}$] | $7.75 \times 10^4$ | $2.77 \times 10^4$ | $7.25 \times 10^4$ | $2.0 \times 10^4$ |

**Table 4.** Planetary input parameters taken from Erkaev *et al.* (2013) for different stellar wind properties and XUV fluxes which correspond to that of the present Sun and an XUV enhancement factor of about 100 times for a higher heating efficiency $\eta$ = 40%.

| Parameter / Case | Earth-type planet, 1 XUV, Case I | Earth-type planet, 100 XUV, Case II | "super-Earth", 1 EUV, Case I | "super-Earth", 100 EUV, Case II |
|---|---|---|---|---|
| Inner boundary: [$R_{pl}$] | 7.84 | 22.94 | 6.98 | 18.44 |
| Inner boundary T [K] | 260 | 4875 | 500 | 4050 |
| Obstacle stand-off distance:[$R_{pl}$] | 8 | 23 | 7.0 | 18.5 |
| Obstacle width: [$R_{pl}$] | 12 | 34.5 | 10.5 | 27.0 |
| Inner boundary density: [$cm^{-3}$] | $7.25 \times 10^4$ | $2.5 \times 10^4$ | $4.0 \times 10^4$ | $1.5 \times 10^4$ |

**Table 5.** Ion escape rates $L_{ion}$ in $s^{-1}$ for a hydrogen-rich Earth-like planet and for a heating efficiency of 15 %, exposed to the XUV flux values which are 1, 10, 50 and 100 times higher than that of the present Sun. The shape of the planetary obstacle is assumed to be similar to an Earth-type magnetosphere.

| H-rich Earth | 1 XUV | 10 XUV | 50 XUV | 100 XUV |
|---|---|---|---|---|
| Case I, $\eta$=15% | $1.35 \times 10^{29}$ | $6.32 \times 10^{29}$ | $1.03 \times 10^{30}$ | $1.59 \times 10^{30}$ |
| Case I, $\eta$=40% | $2.60 \times 10^{29}$ | $1.58 \times 10^{30}$ | $1.40 \times 10^{30}$ | $1.57 \times 10^{30}$ |
| Case II, $\eta$=15% | $2.15 \times 10^{29}$ | $9.50 \times 10^{29}$ | $1.54 \times 10^{30}$ | $2.55 \times 10^{30}$ |
| Case II, $\eta$=40% | $4.11 \times 10^{29}$ | $2.17 \times 10^{30}$ | $2.04 \times 10^{30}$ | $2.40 \times 10^{30}$ |

**Table 6**. Ion escape rates $L_{ion}$ in s$^{-1}$ for a hydrogen-rich "super-Earth" for a heating efficiency of 15 % and 40%, exposed to the XUV flux values which are 1, 10, 50 and 100 times higher than that of the present Sun. The shape of the planetary obstacle is assumed to be similar to an Earth-type magnetosphere.

| H-rich "super Earth" | 1 XUV | 10 XUV | 50 XUV | 100 XUV |
|---|---|---|---|---|
| Case I, $\eta$=15% | $1.03\times10^{25}$ | $4.64\times10^{29}$ | $9.86\times10^{29}$ | $1.66\times10^{30}$ |
| Case I, $\eta$=40% | $8.62\times10^{28}$ | $9.50\times10^{29}$ | $9.71\times10^{29}$ | $1.22\times10^{30}$ |
| Case II, $\eta$=15% | $2.60\times10^{25}$ | $8.18\times10^{29}$ | $1.57\times10^{30}$ | $2.79\times10^{30}$ |
| Case II, $\eta$=40% | $1.62\times10^{29}$ | $1.55\times10^{30}$ | $1.59\times10^{30}$ | $2.80\times10^{30}$ |

**Table 7**. Ion escape rates $L_{ion}$ in s$^{-1}$ for a hydrogen-rich Earth-like planet for a heating efficiency of 15% and 40%, exposed to the XUV flux values which are 1, 10, 50 and 100 times higher compared to that of the present Sun. The shape of the planetary obstacle is assumed to be Venus-like.

| H-rich Earth | 1 XUV | 10 XUV | 50 XUV | 100 XUV |
|---|---|---|---|---|
| Case I, $\eta$=15% | $2.75\times10^{29}$ | $9.02\times10^{29}$ | $1.33\times10^{30}$ | $2.14\times10^{30}$ |
| Case I, $\eta$=40% | $3.72\times10^{29}$ | $1.97\times10^{30}$ | $1.89\times10^{30}$ | $2.54\times10^{30}$ |
| Case II, $\eta$=15% | $5.05\times10^{29}$ | $1.0\times10^{30}$ | $2.23\times10^{30}$ | $2.97\times10^{30}$ |
| Case II, $\eta$=40% | $6.27\times10^{29}$ | $3.23\times10^{30}$ | $3.84\times10^{30}$ | $4.15\times10^{30}$ |

**Table 8**. Ion escape rates $L_{ion}$ in s$^{-1}$ for a hydrogen-rich "super-Earth" for a heating efficiency of 15 % and 40%, exposed to the XUV flux values which are 1, 10, 50 and 100 times higher compared to that of the present Sun. The shape of the planetary obstacle is assumed to be Venus-like.

| H-rich "super Earth" | 1 XUV | 10 XUV | 50 XUV | 100 XUV |
|---|---|---|---|---|
| Case I, $\eta$=15% | $4.81 \times 10^{27}$ | $8.47 \times 10^{29}$ | $1.59 \times 10^{30}$ | $2.74 \times 10^{30}$ |
| Case I, $\eta$=40% | $2.31 \times 10^{29}$ | $1.48 \times 10^{30}$ | $1.69 \times 10^{30}$ | $2.33 \times 10^{30}$ |
| Case II, $\eta$=15% | $1.23 \times 10^{28}$ | $1.38 \times 10^{30}$ | $2.62 \times 10^{30}$ | $4.10 \times 10^{30}$ |
| Case II, $\eta$=40% | $5.17 \times 10^{29}$ | $2.36 \times 10^{30}$ | $2.87 \times 10^{30}$ | $5.36 \times 10^{30}$ |

**Table 9**. Thermal ($L_{th}$) and non-thermal ($L_{ion}$, H$^+$ ion pick-up) loss over the time span of 4.5 Gyr for a H-rich Earth-like planet and a H-rich "super-Earth", orbiting an M dwarf in the habitable zone at 0.24 AU by considering an Earth-type magnetosphere shape for the planetary obstacle form and the low and high heating efficiency of 15% and 40 %.

| H loss: $\Delta t$=4.5 Gyr | H-rich Earth [EO$_H$] | H-rich "super-Earth" [EO$_H$] |
|---|---|---|
| $L_{th}$: $\eta$ = 15 % | ~ 7.0 | ~ 2.2 |
| $L_{th}$: $\eta$ = 40 % | ~ 16 | ~ 8.0 |
| $L_{ion}$: $\eta$ = 15 %, Case I | ~ 0.6 | ~0.41 |
| $L_{ion}$: $\eta$ = 15 %, Case II: | ~ 0.93 | ~ 0.7 |
| $L_{ion}$: $\eta$ = 40 %, Case I | ~ 1.13 | ~ 0.64 |
| $L_{ion}$: $\eta$ = 40 %, Case II | ~ 1.63 | ~ 1.44 |

**Table 10**. Non-thermal ( $L_{ion}$ , $H^+$ ion pick-up) loss over the time span of 4.5 Gyr for an H-rich Earth-like planet and an H-rich "super-Earth", orbiting an M dwarf in the habitable zone at 0.24 AU by considering a Venusian-type planetary obstacle shape and the low and high heating efficiency of 15% and 40 %.

| H loss: $\Delta t$=4.5 Gyr | H-rich Earth [EO$_H$] | H-rich "super-Earth" [EO$_H$] |
|---|---|---|
| $L_{ion}$ : $\eta = 15$ %, Case I | ~ 0.92 | ~ 0.71 |
| $L_{ion}$ : $\eta = 15$ %, Case II | ~ 1.24 | ~ 1.13 |
| $L_{ion}$ : $\eta = 40$ %, Case I | ~ 1.51 | ~ 1.15 |
| $L_{ion}$ : $\eta = 40$ %, Case II | ~2.56 | ~ 2.14 |

**FIGURE CAPTIONS**

**FIG. 1**: Observed hydrogen ENAs by the IBEX spacecraft around Earth. The H ENAs count rate was integrated from 0.7–6 keV on 28 March 2009 from 04:54–15:54 UT. The peak is centered on the subsolar magnetopause and a significant ENA flux extends to ±10 $R_{Earth}$ (Fuselier *et al*., 2010).

**FIG. 2**. Modeling results of Earth's solar wind plasma interaction with the present-day geocorona. Green dots correspond to the solar wind protons, yellow dots represent the neutral hydrogen atoms moving with velocities below 10 km s$^{-1}$ (particles which belong to the atmosphere) while the red and the blue dots represent ENAs with velocities above 10 km s$^{-1}$, moving towards and away from the Sun respectively. The dashed line denotes the magnetosphere obstacle.

**FIG. 3**. Radial profiles for density, velocity and temperature as a function of orbital location in AU and of expected plasma properties of an ordinary stellar wind (solid lines) and during an CME event (dashed lines) on an M-type star with a mass $M_s$ ~0.45 $M_{Sun}$ and a rotation period 2.5 days. For the simulation of stellar wind the initial proton density at 0.1 AU $n_0 = 400$ cm$^{-3}$, temperature $T_0 = 500$ kK, and radial stellar wind velocity $v_{r0} = 300$ km s$^{-1}$ were taken. Simulation of a solar-analogue CME event uses the initial proton density $n_0 = 800$ cm$^{-3}$, proton temperature $T_0 = 1500$ kK, and a radial stellar wind velocity $v_{r0} = 600$ km s$^{-1}$.

**FIG. 4**. Modeled atomic hydrogen coronae and stellar wind plasma interaction around an Earth-like hydrogen-rich planet inside an M star HZ at 0.24 AU (green: protons, yellow: H atoms, blue ENAs flying away from the star, red ENAs flying towards the star; dotted line: magnetopause/planetary obstacle). Fig.4a corresponds to the XUV flux which is equal to that of the present Sun, the moderate stellar wind (Case I) and a lower heating efficiency of 15%. Fig.4b corresponds to the similar input parameters as in Fig.4a, but the XUV flux is 50 times higher. Fig.4c corresponds to the 10 times higher XUV flux than the present one and a heating efficiency of 15%, as well as the moderate stellar wind

(Case I). Fig.4d: corresponds to the similar input parameters as in Fig.4c, but for the Venus-type narrower planetary obstacle.

**FIG. 5**. Modeled atomic hydrogen coronae and stellar wind plasma interaction around a "super-Earth" hydrogen-rich planet inside an M star HZ at 0.24 AU (green: protons, yellow: H atoms, blue ENAs flying away from the star, red ENAs flying towards the star; dotted line: magnetopause/planetary obstacle). Fig5a: the XUV flux is 50 times higher than that of the present Sun, heating efficiency of 15%, the planet is exposed to a moderate stellar wind (Case I). Fig5b: similar conditions except for heating efficiency of 40%. Fig.5c: the XUV flux is 100 times higher compared to that of the present Sun, 40% heating efficiency, moderate stellar wind (Case I). Fig.5d: similar input parameters as in Fig.5c, but exposed to a faster and denser stellar plasma flow (case II).

**FIG. 6**. Illustration of the soft X-ray flux time-dependence for an M dwarf star with 0.4 solar masses (dashed line) and the same curve for a Sun-like star (solid line) in the corresponding HZs normalized by the present Sun flux. The M star within this mass range remains about 200 Myr longer in its activity saturation phase compared to a solar like G star.

# FIGURES

FIG 1

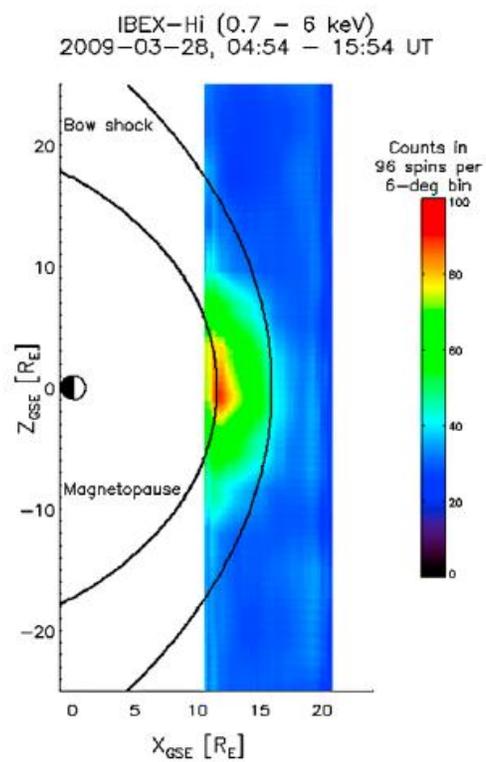

FIG 2

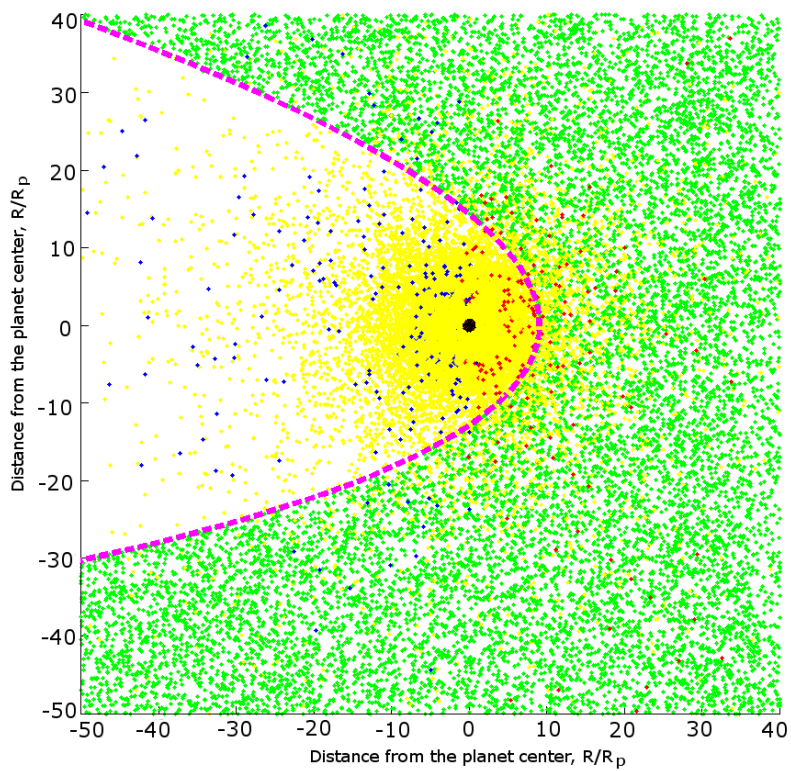

FIG 3

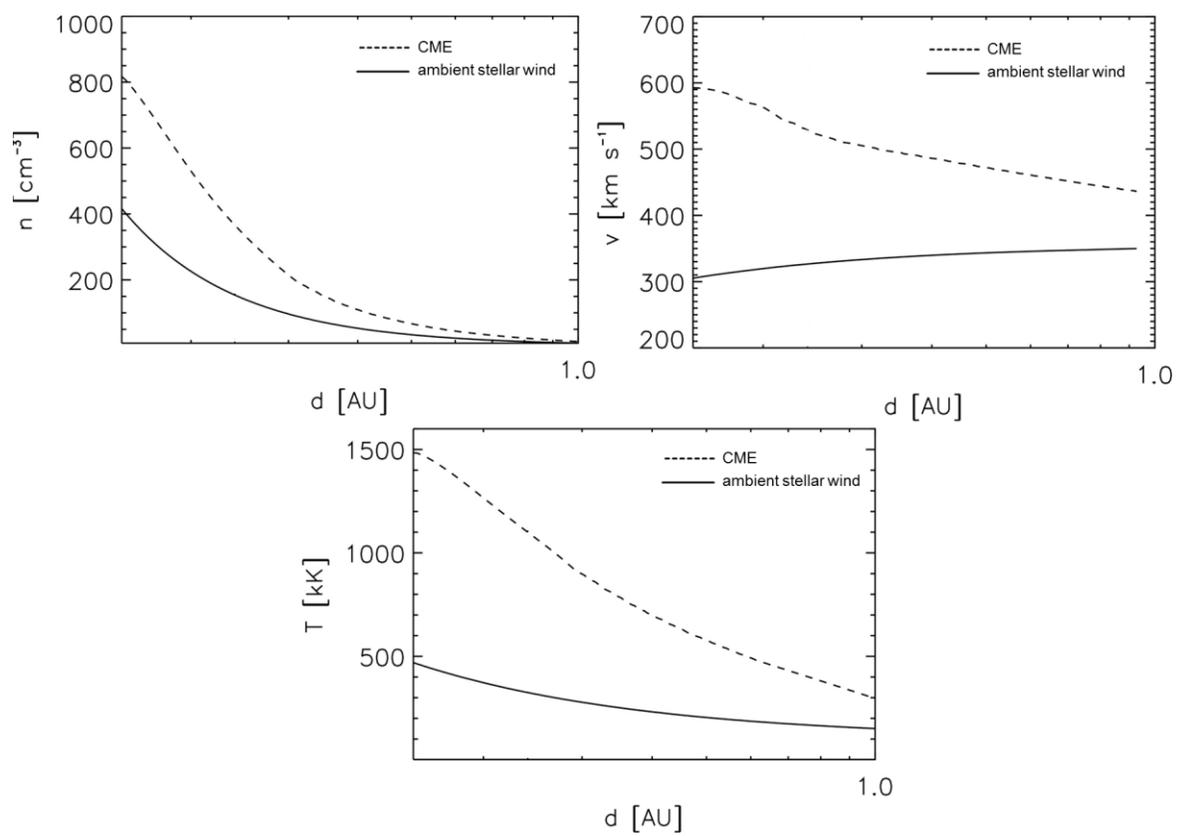

FIG 4

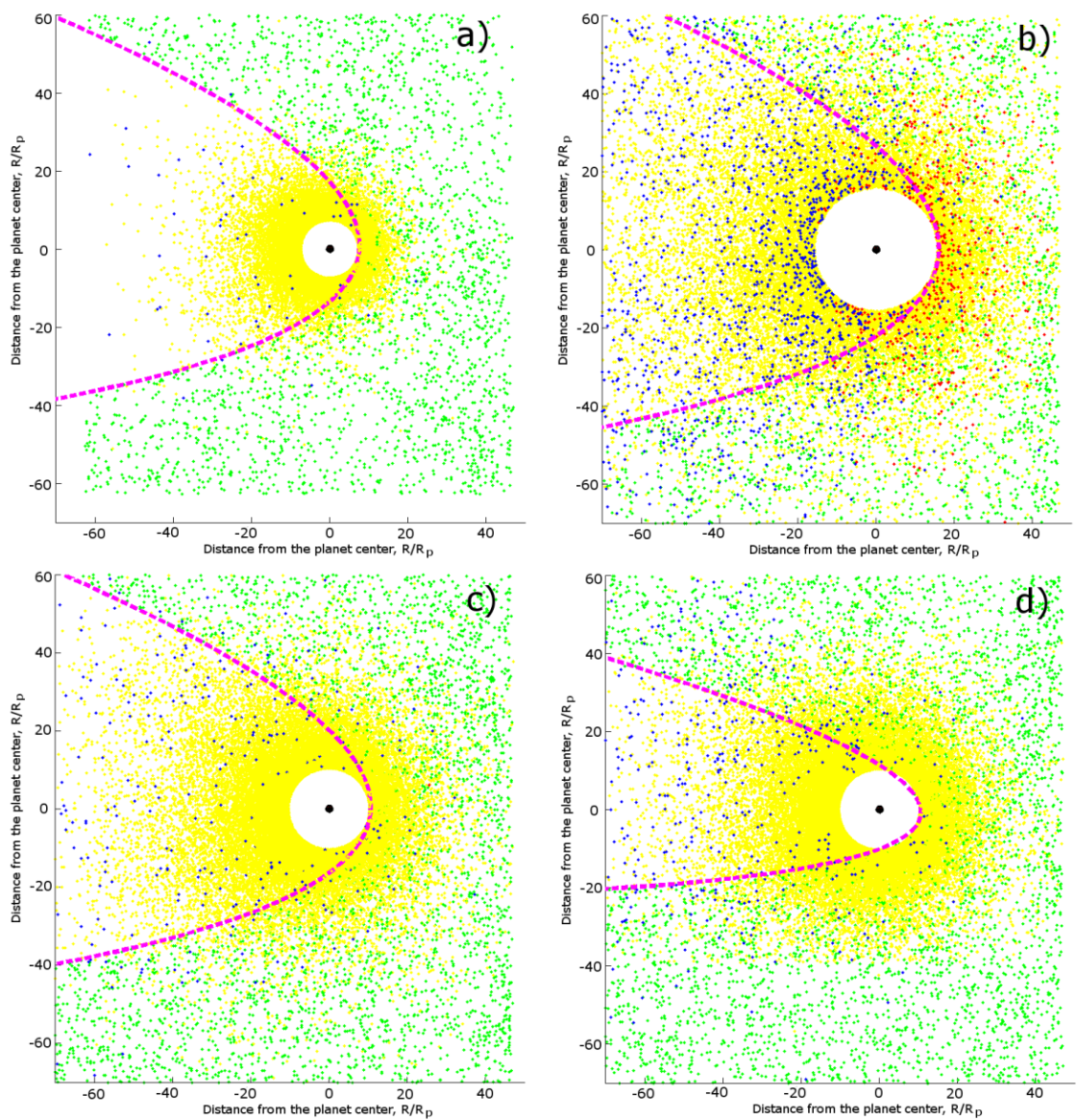

FIG 5

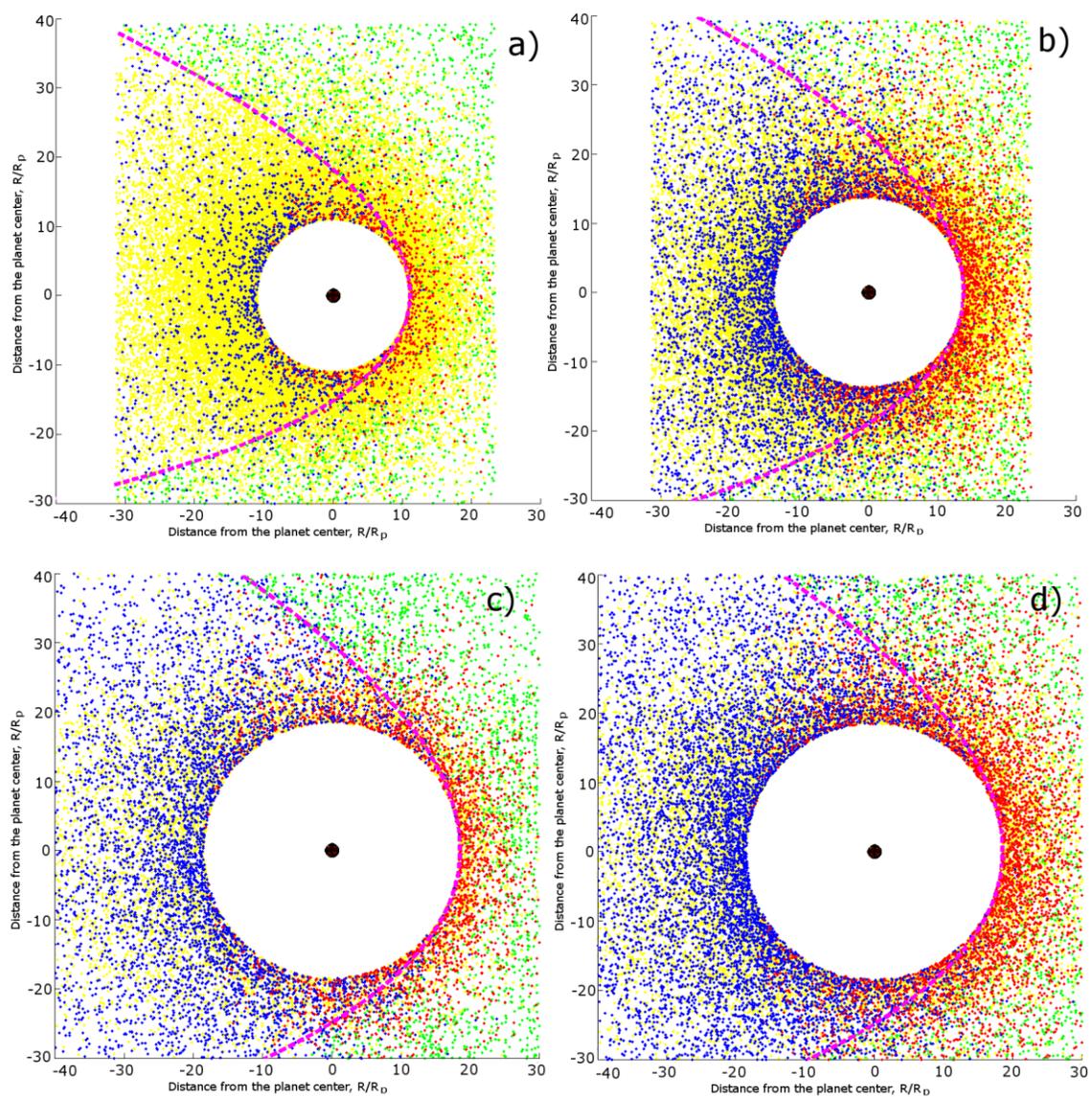

FIG 6

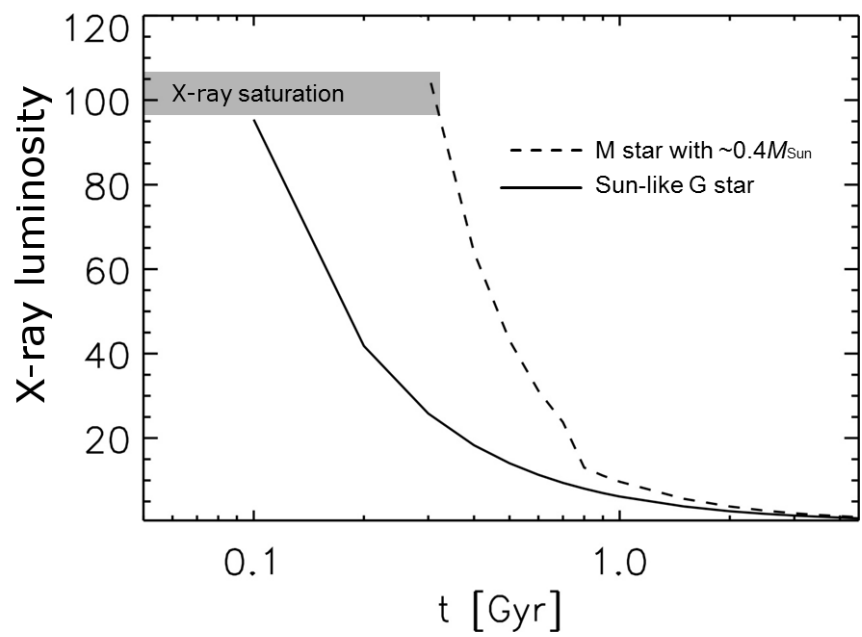